\documentstyle[aps]{revtex}

\title{Chaotic Scattering on a Billiard}
\author{ {Vincent J. Daniels}\\
	{Michel Valli\`eres}\\
	{Jian Min Yuan}\\
          {\small\sl Department of Physics and Atmospheric Science,
                Drexel University,}\\
          {\small\sl Philadelphia Pennsylvania 19104-9984}
        }

\date{\today}

\newcommand{ \beq }{ \begin{equation} }
\newcommand{ \eeq }{ \end{equation} }
\newcommand{ \bea }{ \begin{eqnarray} }
\newcommand{ \eea }{ \end{eqnarray} }
\newcommand{ \penclose}[1]{ \left ( {#1} \right ) }

\newcommand{ \operator}[1]{ {\bf #1} }
\newcommand{ \diff}{ \mbox{d} }

\newcommand{ \derivb}[2]{ {\frac{ \diff #1 }{\diff #2} }}

\newcommand{\sgn}[1]{ {\mbox{sgn}(#1)}}
\newcommand{\cvec}[2]{{\penclose{\begin{array}{c}{#1}\\{#2}\end{array}} } }

\begin{document}
\maketitle

\begin{abstract}
  We investigate  chaotic scattering on an attractive step potential
  with a quadrupolar deformation.    The phase space features of the
  bound billiard  are studied by using the notion of 
  symmetry lines to find periodic orbits. We show that the
  scattering  dynamics is intimately linked to  structures in 
  the bound billiard (infinite potential wall) phase space.  
  The existence of preferred scattering directions  is shown to be a 
  consequence of large scale features of the phase space such as the
  period-two orbits.  Self-similarity in the scattering functions is
  directly linked to unstable periodic orbits of the bound phase
  space.  The main observations and methodology are
  applicable to concave billiards in general.  
\end{abstract}

\section{Introduction}
The study of chaos in billiards has a long and productive history.
Some of the earliest work with billiard systems dates 
back to Bunimovich and his proof that the stadium billiard is ergodic
\cite{Bunimovich74,Bunimovich79}.  Also early on Berry showed that
various deformations of circular billiards  exhibited regular, chaotic or
mixed behavior\cite{Berry81}.  Volumes have been written on various kinds
of closed billiard geometries 
\cite{Sinai70,Robnik83,BroOttGre87,BohBooCarMar93,NocStoCha94,BluAntGeoOttPra96}
as well as billiards with  holes in the wall
\cite{BauBer90,BarJalSto93} to study scattering problems.  However,
little work has been done in studying  
attractive potential scattering with billiard geometries. This
particular type of scattering finds applications in any physical
system that is well described by a discontinuous interface with a
non-spherical geometry such as deformed quantum dots, nuclei, fiber
optics, and semiconductor devices.   In this paper we
discuss the scattering on a quadrupolar deformation of a circle.  We find
that the scattering functions have strongly preferred directions which
persist for a wide range of deformation.  A similar effect is observed
by N\"ockel and Stone \cite{NocSto97} in light scattering in an optical
fiber with a quadrupole cross-section. We show that this behavior is
directly linked to the large scale structure of the phase space which
does not change dramatically as a function of the  deformation.  The
details of the chaotic scattering are related to the presence of
unstable orbits near the critical angle for escape.

We will begin by describing the model system in the context of a
deformed nucleus and introducing the phase space mapping in Section
\ref{sec:model}.    In Section  
\ref{sec:map} we will examine the bound phase space, in particular, we
will use the symmetry lines of the map to determine the symmetric
periodic orbits.  The existence of homoclinic orbits and thus the
chaotic nature of the bound billiard system is shown.  In Section
\ref{sec:scatter} we apply the knowledge of the bound phase space
dynamics to the study of the scattering dynamics using some specific
cases to illustrate general relationships.      In Section
\ref{sec:conclude} we discuss the implications of the connections
described in the paper and the generality of the observations.

\section{Model System}
\label{sec:model}
The model is inspired by the simplest nuclear potential
shape, that is, a simple step potential in two dimensions.  We
introduce an angle-dependent deformation described by  a radial shape
function, $r_s(\theta)$. We are interested in $r_s(\theta)$ that forms
a closed curve in configuration space.  A shape function typically
used to describe  small nuclear deformations can be written 
\beq 
\label{eq:rs1}
	r_s(\theta) = \penclose{1 + b P_2(\sin(\theta))}
\eeq
where $P_2(x)$ is the second Legendre polynomial, all lengths are
scaled by the average nuclear radius, $r_0$, and $b$ is the
deformation parameter.   This shape function
remains reasonable for describing nuclei as long as $|b| < 2/7$.
Beyond this the shape becomes ``peanut-like'', that is, the boundary
becomes partly convex.  For $b = 0$ we have
the circular (undeformed) nuclear step potential. We will focus on
deformations around the middle of this range, $b \approx 0.15$. This
shape defines the location of the step so the  potential is just a
product of the well depth and the step function,
\beq\label{eq:step_def}
        V(r-r_s(\theta)) ~=~ -V_0 S\penclose{-(r-r_s(\theta))}
\eeq	
where $S(t)$ is the unit step function defined by
\beq
	S(t) = \left\{ \begin{array}{lcl}
                     0  & ; & t < 0     \\
                     1  & ; & t > 0
                     \end{array} 
               \right.
\eeq
Figure \ref{fig:potential}
shows a plot of the resulting potential for  $b=0.15$ and $V_0=1.0$.

The Hamiltonian in polar coordinates is  
\beq\label{eq:billiard_hamiltonian}
	H = \frac{p_r^2}{2} + \frac{p_\theta^2}{2r^2} 
		- S\penclose{-(r-r_s(\theta))} 
 		= \epsilon
\eeq  
where $\epsilon$ is the scaled energy, $\epsilon \equiv
\frac{E}{V_0}$, and $V_0$  is the potential depth.  
The potential is a constant within the domain defined by
$r_s(\theta)$; if the total energy is negative, the kinetic energy is
also a constant and the total energy can be scaled away.

The traditional billiard problem is a bound problem with negative
total energy with no possibility of escape.  We generalize this
problem to a scattering situation by  sending particles with positive
total energy from outside the well.  The trajectories followed by
these particles may penetrate the well region, get temporarily trapped
and eventually escape. For this positive total energy case, the total
energy cannot be scaled away.    
The kinetic energy changes as the particle crosses the shape boundary;
 the dynamics at the boundary will depend on the ratio of the kinetic
energy on either side preventing energy scaling.

We  now look briefly at the equations of motion; this is instructive
yet not strictly necessary since we will be developing a map for the system.
  In polar coordinates they are 
\bea 
	\dot{r} & = & p_r \\
	\dot{\theta} & = & \frac{p_\theta}{r^2} \\
	\dot{p}_r & = & \frac{p^2_\theta}{r^3} - \delta(r-r_s(\theta)) \\
	\dot{p}_\theta &=& \frac{\partial r_s}{\partial\theta}
 \delta(r-r_s(\theta)).
\eea 
It is clear from the equations of motion that when $r-r_s(\theta) \ne
0 $ the motion is integrable.  In the event of an intersection with
the wall, $r-r_s(\theta) =0 $, both the radial and angular
momentum change. Zeroes of  $\frac{\partial r_s}{\partial\theta}$
mark  angles for which there is no change in angular momentum.

\subsection{Map}
The dynamics can be rephrased in terms of a mapping.  This follows
from the fact that a particle will deflected only when it
is at the step and undeflected otherwise since the potential is
constant.  This map will relate the coordinates in phase space 
from intersection to intersection with the billiard wall, the
trajectory between collisions being straight lines in configuration space.

There are many different representations for two dimensional maps.  
We will primarily use a map space ($\theta$,$\gamma$) where
 $\theta$ is the angle at which a trajectory intersects the shape
 function.  The second coordinate, $\gamma$, is the angle between the
forward (counter-clockwise) tangent to $r_s(\theta)$ at the
intersection and the incoming momentum vector. The forward tangent is defined by
 $\vec{T}\equiv\derivb{\vec{r}_s(\theta)}{\theta}$.    Other sets of
coordinates that we will use when convenient are ($s,p$) and
($\theta,p$) where $s$ is the arc length measured from $\theta=-\pi$
and $p=\cos(\gamma)$ is the momentum tangential to the shape at an
intersection.    The resulting map in ($\theta,p$) has the form 
\bea
	\theta_{n+1} &=& f(\theta_n, p_n)  \nonumber\\
	p_{n+1} &=& g(\theta_n, p_n).  \nonumber\\
\eea
which we write in terms of the nonlinear map operator, $\operator{M}$,
\beq
	\cvec{\theta_{n+1}}{p_{n+1}} =
		\operator{M}\cvec{\theta_{n}}{p_{n}} 
\eeq
The algorithm for obtaining the map has two steps:
\begin{itemize}
\item{Given the  slope, $m(\theta_n,p_n)$,  and intercept,
 $B(\theta_n,p_n)$,  defining the line along
which the particle moves after the  $n^{th}$  intersection with the
 wall, find the intersection of this  line with the shape function,
 $r_s(\theta)$. This provides  $\theta_{n+1}$.}
\item{Transform the momentum vector $\vec{p}(\theta_n,p_n)$ after the
 $n^{th}$  intersection into the new coordinate system defined at
 $\theta_{n+1}$.  This provides $\gamma_{n+1}$ and therefore $p_{n+1}$.}
\end{itemize}
The first step is the nonlinear part of the map. Generally, this step
involves the solution of 
\beq\label{eq:int_solve}
	r_s(\theta_{n+1})\sin(\theta_{n+1})=m(\theta_n,p_n)r_s(\theta_{n+1})\cos		(\theta_{n+1}) + B(\theta_n,p_n)
\eeq
For $r_s(\theta)$ given by Eq.~\ref{eq:rs1} this is a transcendental
equation and so must be solved numerically.  The second step involves a coordinate system rotation and finding the new 
momentum direction.  The later depends on whether the particle is
transmitted or reflected at the wall. 

 Consider a
trajectory with momentum   $\vec{p}_0$ before an
intersection and $\vec{p}_1$ after.  We decompose these vectors
into components parallel and perpendicular to the tangent at the
intersection.  From the definition of the potential we see that the
force will always be  perpendicular to the shape and pointing to the
inside.  Thus the parallel component of momentum is
conserved so that 
\beq\label{eq:Snell}
        p_0 \cos(\gamma_0) = p_1 \cos(\gamma_1)
\eeq
where $p_0$ and $p_1$ are the magnitudes of the momenta and are
constant everywhere within a given region and 
\[
	p_i=\sqrt{ 2(E-V(r-r_s(\theta))) }.
\]
Equation \ref{eq:Snell} is precisely Snell's law from geometric optics with the
refractive indices being $p_0$ and $p_1$ and the angles measure with
respect to the tangent rather than the normal.  The behavior of a
trajectory intersecting the shape function is  analogous to that of a
light ray striking a polished glass surface.  The obvious difference
being that light can be both 
transmitted {\it and} reflected from the surface while a classical particle
is  transmitted {\it or} reflected.  The rule for transmitting a particle
across the interface is simply that if it can be transmitted it will
be, otherwise it will be reflected.  More precisely, if
\beq\label{eq:trans_cond}
	|\cos(\gamma_0)| < p_1/p_0 
\eeq 
the particle is transmitted with the angle $\gamma_1$  given by
Eq.~\ref{eq:Snell}. In analogy with total internal reflection in
optics, $\gamma_1 = \pi/2$ leads to
\beq\label{eq:crit_cond}
	|\cos(\gamma_0)| = p_1/p_0.
\eeq
The solutions of Eq.~\ref{eq:crit_cond}, denoted by $\gamma^+_{cr}$
and $\gamma^-_{cr}$, are 
centered on $\gamma=\pi/2$.  Particles incident within  
\beq\label{eq:crit_range}
\gamma^-_{cr} < \gamma < \gamma^+_{cr} 
\eeq
cross the billiard boundary. 

A particle not  transmitted across the
boundary  must stay in the region it came from.   
Since the parallel component of momentum is still conserved,
Eq.~\ref{eq:Snell} still holds with  $p_0=p_1$ yielding 
\beq\label{eq:a}
        \cos{\gamma_0} = \cos{\gamma_1}.
\eeq
This embodies the law for specular reflection which in our coordinate
system must be taken to be $\gamma_1 = -\gamma_0$, that is reflection
by the same angle with respect to the tangent.
This is the only possibility for the bound problem, namely, $E<0$.

Applying these rules to scattering billiards leads to the following
results. A particle  incident from the outside (asymptotic
region)  will always be transmitted to the inside (interaction region).
On the outside we have $p_0=\sqrt{2E}$ while on the inside
$p_1=\sqrt{2E+2}$ so 
\beq
	p_1/p_0=\sqrt{1+1/E} > 1 > |\cos{\gamma_0}| ~\forall~ E>0.
\eeq
 On the other hand, for a particle incident from the inside
\beq\label{eq:trans_refl}
	p_1/p_0 = \sqrt{E/(E+1)}= |\cos(\gamma^\pm_{cr})| < 1 ~\forall~ E>0.
\eeq
So the  particle  may be transmitted $(|\cos(\gamma_0)|\leq\sqrt{E/(E+1)})$
or reflected $(|\cos(\gamma_0)|>\sqrt{E/(E+1)})$. Equation
\ref{eq:trans_refl} shows that there is always a critical angle for 
reflection no matter how energetic the particle is.   

From the scattering perspective we are only
interested in trajectories that will intersect the shape function.
These interesting scattering initial conditions  will be in one to one
correspondence to points in the bound map.  When a scattering
trajectory is exiting the 
interaction region the map point will have a negative $\gamma$
value.  However, there is a one to one mapping of the negative $\gamma$
values to positive $\gamma$ values given by Eq.~\ref{eq:Snell}.  These
values will correspond to the $\gamma$ value just before the escape
intersection or, equivalently, to the $\gamma$ that would have resulted
if the particle did not escape.  The later mapping for exiting
particles is the one that we will use.  With this prescription  both
the bound and the scattering map have the same domain, namely, $0 \leq
\gamma \leq \pi $ and $ -\pi < \theta \leq \pi$.  The difference is
that for the scattering map the phase space is partitioned into an
asymptotic region between $\gamma^-_{cr}$ and $\gamma^+_{cr}$ and the
interaction or bound region which makes up the rest of the phase
space.

\section{Bound Map and Phase Space}
\label{sec:map} 
The bound and
scattering maps share the same phase  space;  it follows that the
scattering problem is  influenced by the same structures in phase
space as that of the bound map. With this in mind we first study the
bound problem.

\subsection{Periodic Orbits and Symmetry Lines}
Knowledge of the periodic orbits is key to understanding the phase
space of the map.
  Several orbits and their map coordinates can
be guessed quite easily; for instance, the period two (P2) orbits,
one along the $x$-axis and the  
other along the $y$-axis.  While the existence of others may be obvious,
their coordinates are not easy to find; like the period three (P3) and
four (P4) orbits. There are still others whose mere existence may not be as
obvious, i.e., those generated via bifurcations of the above
mentioned orbits.  

An elegant construct for finding periodic orbits in ``reversible''
maps involves the use of {\it symmetry
lines}\cite{Mackay82,PinLar86,Heagy89,HeaYua90}. These follow from the
fundamental symmetries of the problem; in our case time reversal
invariance and geometrical symmetries of $r_s(\theta)$.  
This leads to three distinct symmetry operations for our map: momentum
reversal ($\operator{R_p}$), reflection about the $x$-axis ($\operator{R_x}$), and reflection about the
$y$-axis ($\operator{R_y}$).  They are defined by 
\bea\label{eq:rp_sym}
	\operator{R_p} \cvec{\theta}{p} &=& \cvec{\theta}{-p} \\
\label{eq:rx_sym}
	\operator{R_x} \cvec{\theta}{p} &=& \cvec{-\theta}{p} \\
\label{eq:ry_sym}
	\operator{R_y} \cvec{\theta}{p} &=& \cvec{\sgn\theta\pi-\theta}{ p}
\eea
with 
\[
 \sgn{\theta}
 \equiv \left \{ \begin{array}{l l} 1 &;~~ \theta \ge 0 \\ -1 &;~~
 \theta < 0\end{array} \right . 
\]
The operators $\operator{R_p},\operator{R_x},\operator{R_y}$ are
clearly their own inverses.  Each also inverts the map, 
\beq\label{eq:invert}
	\operator{M}^{-1} = \operator{TMT}
\eeq
where $\operator{T}$ represents any one of the three operators,
$\operator{R_p},\operator{R_x}$, or $\operator{R_y}$ 

All reversible area preserving  maps can be factored into a product of
two {\it orientation reversing involutions} or symmetry
operators\cite{Mackay82}, 
\beq
	\operator{M} = \operator{I_1}\operator{I_0}.
\eeq
We are interested in the three factorizations of the map given by: 
\beq\label{eq:factors}
	\operator{M} = \operator{I_1}\operator{I_0} = \{\operator{MT}\}
	\operator{T}.
\eeq
So 
\bea
	\operator{I_0} &=& \operator{T} \\
	\operator{I_1} &=& \operator{MT}.
\eea
The above operators represent the first two in an infinite hierarchy of
symmetries defined by 
\beq\label{eq:sym_A0}
	\operator{I}_n = \operator{M}^n\operator{I_0},~~~~n = 0,1,. . . 
\eeq
Where $\operator{M}^n$ represents $n$ compositions of $\operator{M}$.

To utilize this infinite hierarchy of symmetries we need to determine
the invariant sets or {\it symmetry lines} associated with each of the
involutions, $I_n$.   The symmetry lines are the solutions of 
\beq\label{eq:sym_B0}
	\Gamma_n : \{ \xi |\operator{I_n}\xi = \xi\}
\eeq
where $\xi \equiv (\theta,p)$ is a point in the map.  The symmetry
lines satisfy the following recursion relation \cite{PinLar86}
\beq
\label{eq:sym_B1}
	\operator{M^m}\Gamma_n = \Gamma_{2m+n}.
\eeq
Letting $n=0$ will generate all of the {\it even} symmetry lines:
\[
	\operator{M^m}\Gamma_0 = \Gamma_{2m}.
\]  
Letting $n=1$ will generate all of the {\it odd} symmetry lines:
\[
	\operator{M^m}\Gamma_1 = \Gamma_{2m+1}.
\]   
We will call the  lines $\Gamma_0$ and $\Gamma_1$ the {\it fundamental
symmetry lines} of the map.  The fundamental symmetry lines for each
of the possible involution pairs can be calculated analytically and
are presented in Table \ref{tbl:gamma}.

Note that there are two branches for each of the spatial symmetry
lines while there is only one branch for the $\Gamma^p_0$ symmetry
line and no solution for $\Gamma^p_1$.  The immediate result of the
non-existent $\Gamma^p_1$ is that there are no odd symmetry lines for
momentum reversal.  It follows  that there are no momentum reversal
symmetric odd period orbits. 

A periodic orbit is defined by
\beq\label{eq:per_orbit}
	P_n ~:~ \{ \xi | M^n\xi=\xi\},
\eeq
that is, the $n$ period points are invariant under $n$ applications of
the map. This set includes orbits whose periods are divisors of $n$ as
well.  The important result from the symmetry line theory
\cite{Heagy89,Mackay82} is
\beq\label{eq:sym_inter}
	\Gamma_n \cap \Gamma_m \subset P_{n-m}.
\eeq
Using this method the search for symmetric periodic orbits reduces to
finding the intersections between different symmetry lines.  

Figure
\ref{fig:sym_lines} shows the symmetry lines from $\Gamma^x_0$ to
 $\Gamma^x_9$ for the deformation, $b=0.15$.   Each of these lines has
two branches; the two vertical lines at $\theta=0$ and $\theta=\pi$
are  $\Gamma^x_{0,0}$ and $\Gamma^x_{0,1}$ and the two branches $0$
and $1$ of $\Gamma^x_{1}$ to $\Gamma^x_{9}$ start at ($0,\pi$) and
($\pi,\pi$) respectively. Periodic orbits can be obtained from the
intersections of these symmetry lines. As an example Fig.~\ref{fig:sym_x_0_9}
shows the intersection  $\Gamma^x_0 \cap \Gamma^x_9$ (We drop the $x$
superscript label from here on). 

The periodic orbits P3 and P9 are located at the intersections of
these two lines.   The orbits are labeled by their winding numbers
which is determined by counting the number of times, $k$, the symmetry
lines have ``wrapped around'' the cylinder defined by $\theta$
beginning with the smallest $\gamma$.  We
will come back to the fact that the orbits labeled $6/9$ and $3/9$
consist of three intersections each.  The
winding number of the orbits obtained from $\Gamma_{n,j} \cap \Gamma_{m,i}$
is formally given by   
\beq
	\omega = \left\{ \begin{array}{l l}
		\frac{ 2 |k_m-k_n|}{|m-n|} &;~~i = j \\
		\frac{ 2 |k_m-k_n| +1}{|m-n|} &;~~i \neq j 
		\end{array} \right\} ; k_m = 0,. . .,|m|-1~;~k_n = 0,
. . ., |n|-1,
\eeq
where $k_m$ ($k_n$) is the number of times the
 $\Gamma_{m,i}$ ($\Gamma_{n,j}$) symmetry line has wrapped around the
cylinder.  As shown in Fig.~\ref{fig:sym_x_0_9} the
 $\Gamma_{0,0}\cap\Gamma_{9,0}$ and $\Gamma_{0,0}\cap\Gamma_{9,1}$ gives
us all of the P9 orbits that have a vertex at $\theta=0$ and 
reflection symmetry about the $x$-axis.  All of the P9 orbits labeled
in Fig.~\ref{fig:sym_x_0_9} are shown in Fig.~\ref{fig:p9_orbits}.
The $\Gamma_{0,1}\cap\Gamma_{9,0}$ and $\Gamma_{0,1}\cap\Gamma_{9,1}$ give
all of the P9 orbits that have a vertex at $\theta=\pi$ and
reflection symmetry about the $x$-axis.   

\subsection{Bifurcations}
Bifurcation diagrams summarize the behavior of dynamical systems as
parameters are varied; they describe how periodic orbits are created,
destroyed or collide. The symmetry lines can be used to obtain
bifurcation information.  In Fig.~\ref{fig:p9_orbits} there are three
orbits labeled with winding number $6/9$ (and three with $3/9$).  The
local twisting of the $\Gamma_9$ symmetry line indicates that a
bifurcation of some kind has occurred at some smaller $b$ value.  In
Fig.~\ref{fig:p9_orbits} the sequence of orbits on the right show the
three orbits resulting from this twist of the symmetry line.  The ordering
is as shown: the $2/3$ orbit is between the two P9 orbits labeled $6/9+$
and $6/9-$. We might naively suspect that this is the
result of some kind of period tripling bifurcation; however, after a
closer look we see that this is not the case.  It is instructive to 
examine this segment of the symmetry lines as a function of $b$.

Figure \ref{fig:p9_bif_close} shows a blow up series of the
$\Gamma_{9,0} \cap \Gamma_{0,0}$ for different values of $b$ ranging
from before to after the bifurcation.  This series reveals that before
the bifurcation there is only the $2/3$ orbit (e.g., $b_1 = 0.12175$).
At $b=b_2=0.121787$ there is a saddle center bifurcation that creates
a stable ($6/9+$) and unstable ($6/9-$) P9 orbits.  After this
bifurcation the two P9s are next to each other and the P3 is below
them (e.g., $b_3=0.1218$) until the  unstable P9 collides   
with the P3 orbit at $b=b_4=0.12187$ and they pass through  each other
resulting in the situation pictured in Fig.~\ref{fig:p9_orbits}.  Figure
\ref{fig:p9_bif_phase} shows the phase space around one of the P9 
fixed points (a) after the tangent bifurcation but before the
 collision ($b=b_3$) and (b) after the collision ($b=b_5=0.12195$).
This type of bifurcation is referred to by 
MacKay\cite{Mackay82,MaoDel92} as an  ``m bifurcation'' and is a generic
bifurcation of area preserving maps. The bifurcation diagram for this
bifurcation is shown in Fig.~\ref{fig:m_bif}.  The original orbit is
labeled by  its  winding number, $6/9$. Two P9 orbits are created in a
saddle center and the unstable one passes through the original period
$6/9$ orbit. The two steps to this bifurcation are: the saddle center
creation of two orbits followed by the collision of one of these with
the generating orbit. These two steps occur at different values of
 $p=\cos(\gamma)$; in general, they may occur at the same value of $p$,
$\Delta p = 0$.  All of the generic bifurcations of area
preserving maps are discussed by MacKay\cite{Mackay82}. 

\subsection{The P2 and P4 orbits}
The current investigation is primarily focused on a billiard with
a deformation of $b=0.15$.  
At $b=0$ the billiard is circular and the phase space consists of
periodic orbits with rational winding numbers (resonant tori) and
quasiperiodic orbits with irrational winding number (irrational
tori).  As the $b$ is increased from zero the resonant tori are all
destroyed leaving isolated periodic orbits.  On the other hand, many
of the irrational tori persist for $b>0$.  These irrational tori
stretch across the phase space creating natural momentum boundaries.  At
 $b=0.15$ nearly all of the original irrational tori have been
broken leaving nothing but isolated islands 
and chaotic orbits.  So, in principle there is no dynamical
partitioning of the phase space into different momentum regions as is
the case with smaller deformations.  Thus the phase space is expected
to be quite complicated.  Yet, in spite of this complexity  the low
period orbits play a key role in the dynamics.   

Of particular importance are the P2 and P4 orbits.  There are two P2
orbits one stable and the other unstable.  The stable orbit lies along
the $x$-axis and has the map coordinates ($0,\pi/2$),($\pi,\pi/2$);
the two periodic points are surrounded by large regions filled with
KAM tori.  Figure \ref{fig:tori} shows the KAM regions associated with
the stable P2 orbit along with several other stable periodic orbits
and two chaotic trajectories.  Also seen in this Fig.~\ref{fig:tori}
are the stable P4 orbits with winding number $1/4$ and $3/4$. Their
tori are surrounded by four P8 orbits creating during an
``m-bifurcation'' of the P4 orbits.  In configuration space the
stable P4 is diamond shaped with vertices at $\theta=0$, $\pm\pi/2$, and
 $\pi$.   

The unstable counterparts of the P2 and P4 orbits mentioned above are
equally important.  The unstable P2 orbit lies along the
 $y$-axis and has the map coordinates
($\pi/2,\pi/2$),($-\pi/2,\pi/2$).  The unstable P4 orbits  trace out a
rectangle in configuration space, with  the long 
sides parallel to the $y$-axis.  The $1/4$ orbit follows this path in a
counter-clock-wise direction while the $3/4$ orbit follows the same path in a
clock-wise direction.  Unstable orbits are characterized by their
stable and unstable  manifolds.  Figure \ref{fig:2_4mans} shows
approximations to the stable manifolds of the unstable P2 orbit and
the P4 orbit with winding number $3/4$.   A manifold is an invariant
set under the  mapping (or inverse mapping); in other words, a point
on a manifold maps to a point  on a manifold.  Therefore, given any
subset of the manifold we can generate the whole manifold.  The simplest
(approximate) subset of the manifold can be obtained by linearizing
the map around a periodic point.  The linearized map provides the
``stability matrix'' which, for simple closed billiards, has a general
analytical solution given by Berry\cite{Berry81}.  The stability
matrix  can be diagonalized to find the stable (negative eigenvalue's
eigenvector) and unstable (positive eigenvalue's eigenvector)
directions at the periodic points.  To generate the stable manifolds
shown in Fig.~\ref{fig:2_4mans} we begin with a large ($10^5$) set of
initial conditions from a 
small segment ($10^{-8}$) of a line lying along the stable direction and
centered on a periodic point. These initial conditions are then
iterated under the inverse map, $\operator{M^{-1}}$, for about $30$
iterates.  Beyond $\sim 30$ iterations small deviations  of the
initial set of points from the actual manifold start to become large
producing large deviations from the actual manifold.

The P2 and P4 orbits' manifolds and KAM regions occupy a large portion
of the map and so are important to understanding the dynamics.
However, there are an infinite number of other periodic orbits which
play a role in the dynamics.  The ones that are particularly important
for the analysis in the next section are periodic orbits whose stable
islands have bifurcated away leaving nothing but unstable periodic
points and their manifolds. A particular set of such orbits which we
we will examine more closely are those  with winding numbers between
$1/2$ and $3/4$.  The manifolds of these orbits are sandwiched between
and intimately intertwined with the P2 and P4 manifolds of
Fig.~\ref{fig:2_4mans}. In the next section we will relate the gross scattering
properties to the P2 orbits and the finer scale structure of the
scattering to the low period unstable orbits near the asymptotic
region of the scattering phase space. 

\section{Scattering}
\label{sec:scatter}	
 A scattering experiment consists of launching positive energy particles from an
asymptotic region  towards an interaction region and observing their
``states''  upon leaving the interaction region.  For the billiard problem the
asymptotic region is the area outside the billiard  and the 
interaction region is the area inside and including the billiard
boundary.   The ``state'' observed on leaving the interaction is
often the escape angle, $\Phi$, that a trajectory makes to an arbitrary fixed
axis.  Another interesting quantity to observe is the ``delay time''
which is defined as the amount of time a particle remains in the
interaction region.   

In terms of the bound billiard map the 
asymptotic region is defined by the area between the critical angles,
 $\gamma^\pm_{cr}$.  The interaction region is the area above
 $\gamma^+_{cr}$ and below $\gamma^-_{cr}$.  Figures \ref{fig:tori}
and \ref{fig:2_4mans} each show the bound phase space with the
critical angle lines  $\gamma^\pm_{cr}$ separating the asymptotic from
the interaction region.  A scattering trajectory
will have exactly two map points in the asymptotic region (between
 $\gamma^\pm_{cr}$); one corresponding to its entry into the billiard
and the other corresponding to its exit.  In general a scattering
trajectory may also have an arbitrary number of points in the
interaction region of the map corresponding to being trapped in the
potential region.  The number of map points, $n$, that a trajectory has in the
interaction region corresponds to the number of times it hits the
billiard wall without escaping (bounces); this is effectively
equivalent to the delay time.  Thus the scattering map
looks just like the bound map with horizontal lines corresponding to
the critical angles dividing the phase space into asymptotic and
interaction regions.     

In scattering problems one typically defines an impact parameter
which is a simple  function of initial conditions.  For  two
dimensional non-integrable dynamical systems  a general analysis
requires that the space of impact parameters also be two-dimensional.
However, as we will demonstrate later, a well chosen one-dimensional
impact parameter is sufficient to characterize the chaotic nature of
our billiard scattering system.   For our analysis we launch particles
from a line at fixed $x_0$ outside the well and parallel to the $y$-axis
with fixed momentum components ${p_x}_0 = \sqrt{2E}$, and ${p_y}_0=0$.  The
impact parameter is  $y_0$.  We will record the scattering functions
 $\Phi$ and $n$ as functions of $y_0$.   The angle $\Phi$ is measured
with respect to the $+x$-axis.    Figure 
\ref{fig:fract_2} shows a series of enlargements of the scattering
function $\Phi(y_0)$ for $b=0.15$.    The level-0
plot excludes the relatively uninteresting range of impact parameters
$0< y_0<1.1$ where trajectories bounce only one, two or three times
before escaping.  Note that in the range of impact 
parameters shown all trajectories  bounce  four or more 
times before escaping.   There 
are several interesting features of these plots.  First, there 
are clearly regions where the scattering is regular, i.e., piece-wise
continuous, separated by ``unresolved''  regions.  Second, the
``odd'' enlargements (e.g., levels-1 \& 3) are mirror images of the ``even''
enlargements.  Third, the scattering is  predominately in the forward
and backward directions  while other angles are clearly excluded
indicating preferred scattering directions. Fourth, the figures show a
striking self-similarity which persists on all scales attainable with
double-precision floating point arithmetic. We will address each of
these points in turn and then ask whether such features persist when
we allow all possible scattering initial conditions. 

The behavior of the scattering function of Fig.~\ref{fig:fract_2} is
the hallmark of a chaotic scattering system
\cite{EckJun86,JunSch87,BleOttGre89,DanValYua93}. The smooth regions are sets
of initial conditions that bounce the same number of times before
exiting.  They are separated by regions where the escape angle
appears unresolved or ``chaotic'' as a function of impact parameter.  As
Fig.~\ref{fig:fract_2} shows this behavior persists to higher
magnification of an unresolved region so that at any magnification
there are always chaotic regions in the escape angle function.
Figure \ref{fig:fract_1} shows (top) the number of bounces, $n$, before
escaping versus the impact parameter and (bottom) the escape angle
versus impact parameter (this is \ref{fig:fract_2}a).   
This shows that the chaotic  regions are related to trajectories that
bounce a large number of times before escaping.  The more bounces a
particle undergoes the more  sensitive the outgoing angle is to small 
displacements in the impact parameter.  The singularities in the
function $n(y_0)$ form an uncountable infinity of points; the impact
parameters that lead to infinite $n$ form a fractal set. 

The origin of this fractal set lies in the stable manifolds of
the homoclinic orbits of the system.  The manifolds themselves are 
of measure zero in  phase space so that a typical initial condition
will not fall exactly on a manifold.  However, initial conditions that find
themselves near one of these manifolds will tend to move 
toward  the periodic orbit before moving away.  If the periodic orbit
is entirely in 
the interaction region then the particle may be trapped for an
arbitrarily long time as it approaches the periodic orbit.  For
chaotic scattering to occur these manifolds must reach into the
asymptotic region where the initial 
conditions live.  For billiard systems the asymptotic region is
defined by the critical angles for escape.  Thus there can be chaotic
scattering only if the stable manifolds of periodic orbits which live in the
interaction region cross the critical angle.

\subsection{P2 and P4 Manifolds}
Figure \ref{fig:2_4mans} shows an approximation to the stable
manifolds of the P2 orbit and one of 
the P4 orbits along with the critical angle lines defining the
asymptotic region.  We also show the line of initial conditions, $I$,
that produced the results in Figs. \ref{fig:fract_2} and \ref{fig:fract_1}.
The P4 manifold lays on the P2 manifold and mixes with it at their
``interface''.  In between the P4 and P2 manifolds there are an
infinite number of other unstable periodic orbits whose stable
counterpart periodic orbits have bifurcated at smaller deformations
and whose stable manifolds are intimately 
intertwined with the P2 and P4 manifolds. As we will see later, these
orbits are responsible for the structure  of the scattering functions,
in particular, the scaling and relative sizes of smooth regions seen in
the scattering functions.  

Scattering trajectories that are initialized near the stable
manifolds of periodic orbits living in the asymptotic region will
have very short scattering times.  For example, the unstable P2 orbit
is such an orbit.   Its stable manifold dominates the asymptotic
region  so that most initial conditions will find themselves close to
it; the resulting trajectories will bounce only a few times, if they
bounce at all,  before  escaping. Scattering trajectories that are initialized
near the stable manifolds of periodic orbits that live in the
interaction region will exhibit long scattering times.  The P4 orbit
is an example of this type; its stable manifold reaches the asymptotic region
where scattering trajectories may come close to it.

\subsection{Preferred Scattering Directions}
The presence of the large KAM zones around the stable P2 orbit 
restricts the manifolds (and therefore, the chaotic trajectories) to
the two   ``neck'' regions around  $\theta=\pm\pi/2$. It is ultimately
a result of the existence of large P2 KAM zones in the phase space
that restricts particles to exit in either one of these neck regions.
The extent of the P2 manifolds limits the momentum range of the
escaping particles. Thus the P2 orbits are  the source of the
bidirectional nature of the scattering functions.  In this section we
present a way to get a more quantitative measure of the 
degree of directionality found in the escape angle function. 

The preferred directions apparent in Fig.~\ref{fig:fract_2} are
independent of the particular choice of scattering initial
conditions. That is, the same preferred directions and range of escape
angles results from nearly any line of initial conditions that we choose. 
To show this consider a set  of initial conditions consisting of the two lines
 $\gamma^+_{cr}$ and $\gamma^-_{cr}$ under one forward iteration of
the map.  Iterating the boundary between the asymptotic and
interaction regions in this way we obtain an image of the boundary
one bounce later.  Since the mapping is orientation preserving the
phase space above $\gamma^+_{cr}$ maps to the phase space above
 $\operator{M}\gamma^+_{cr}$ and  below $\gamma^-_{cr}$ maps to 
below $\operator{M}\gamma^-_{cr}$.  Figure \ref{fig:scan_b1}
clearly shows  there are well defined areas of the asymptotic region  
into which trapped particles must go. The allowed regions of escape
for particles that have bounced at least  once are defined by the four
shaded lobes in Fig.~\ref{fig:scan_b1}. It also shows that at this 
energy ($E=0.2856$) and deformation ($b=0.15$) particles that are
trapped in the upper phase  space (clockwise rotating orbits) will
never reach the lower phase space (counterclockwise rotating orbits).
 
The areas defined by the shaded lobes of Fig.~\ref{fig:scan_b1} are
easily converted into escape angle ranges.  The 
plot shown in Fig.~\ref{fig:sketch} shows these escape angle ranges
overlaid on the billiard shape for initial conditions above
 $\gamma^+_{cr}$, that is, clockwise rotating orbits.  We find that
the escape angle must fall within the two limits given by 
\bea\label{eq:ranges}
	-1.0709 <& \Phi <& 0.9355 \nonumber\\
	-4.2125 <& \Phi <& -2.0609.
\eea 
These limits apply to all initial conditions starting anywhere in the upper
interaction region.  The range of escape angles observed in 
Fig.~\ref{fig:fract_2} certainly fall within these bounds. Later we
will address the fact that the escape angle ranges observed in
Fig.~\ref{fig:fract_2} are much
smaller than the ones given in Eq.~\ref{eq:ranges}.  Thus, we have
an upper bound on the range of escape angles for any set of initial
conditions in the upper interaction region. The ranges can be reflected
about $\Phi = 0$ to obtain the range for initial conditions in the
lower interaction region.

We can verify the existence of the preferred directions  as well as
add distribution information by launching a large set of initial
conditions evenly distributed in the  asymptotic region.  We consider
 $10^6$ initial conditions in the  rectangle  given by $-\pi < \theta <
\pi$ and $\pi/2 < \gamma < \gamma^+_{cr}=2.061596$ (corresponding to
E=0.2856) keeping track of the number of bounces before
exiting. Figure  \ref{fig:with_fly_by} shows a histogram of the number
of trajectories, $N$, binned according to (a) escape angle,  $\Phi$,
and (b) the $\theta$ value at escape, $\Theta$, for all trajectories on
exiting, including ``fly by'' (no bounce) trajectories.   
We see that even when  all trajectories are included there are
strongly preferred scattering  directions.  In
Figs. \ref{fig:no_fly_by}(a-b) we show the 
same data with the fly by trajectories removed.  
This now reproduces the escape angle ranges given by
Eq.~\ref{eq:ranges} and indicates that most of the flux is  towards 
the right half of the allowed range.  If we consider excluding
trajectories that have bounced twice as well we find that the escape
angle range is still over estimated.  Finally, Fig.~\ref{fig:3_plus} shows only
trajectories which have bounced four or more times (there are no three
or five bounce trajectories). Here we see that the range of escape
angles $\Phi$ is precisely what is observed in Fig.~\ref{fig:fract_2}.
If we continue the process of excluding the fewest bounces further we
will find the distribution and range remains essentially unchanged. 
This rapid convergence with the number of bounces implies that it
takes about four bounces for a chaotic scattering trajectory to
``forget'' where it came from.  This 
is so since the distribution of outcomes of all possible initial
conditions is well reproduced by a  small sample of initial conditions. Of
course the small sample of initial conditions cannot be completely
arbitrary; it must pass through a ``relevant'' region of phase space.
What, precisely constitutes the relevant region is the subject of the
next paragraph.   

\subsection{Chaotic Region}\label{sec:chaotic}
We can invert the question answered by Fig.~\ref{fig:scan_b1} and ask:
``What part of the asymptotic region leads to trajectories that bounce
at least once?''  To answer this we take the critical angle lines and
iterate them once with the inverse map to get Fig
\ref{fig:scan_b2}. Trajectories beginning in the shaded lobes will
enter the interaction region for at least one bounce. Consider this
picture overlaid on Fig.~\ref{fig:2_4mans}.  A large portion 
the shaded region of Fig.~\ref{fig:scan_b2} is occupied by the stable
manifold of the P2 orbit.  This portion of the shaded regions will
lead to nearly all of the one and two bounce orbits.  The manifolds
that lie between the P2 and P4 are responsible for the longer orbits. 
Most important are unstable periodic orbits whose periodic points lie
entirely in the interaction region but whose manifolds reach into the
asymptotic region.  These are responsible for the structure of the
scattering functions.  

To illuminate this connection we look at the escape time functions
associated with the series of enlargements of Fig.~\ref{fig:fract_2}.
These are shown in Fig.~\ref{fig:fract_3}.  We choose to enlarge the
chaotic regions between the two largest smooth regions inside a
chaotic band; thus we are
looking at the largest scale feature of the scattering functions.
Clearly one level can be scaled into another.  We have already identified
the $n$-scale by offsetting each enlargement by seven.  In
Fig.~\ref{fig:fract_3a}  we have plotted the trajectories associated
with the impact parameter at the center of the largest smooth regions
in each level, i.e., $n=6$ of level 0 in Fig.~\ref{fig:fract_3}.  We
see that these trajectories all start near the period $5/7$ orbit (shown in
Fig.~\ref{fig:p7_orbit}) and follow it around one more time for each
enlargement.  This implies that the period seven orbit should have a periodic
point very close to asymptotic boundary and, 
since the trajectory can bounce at all seven vertices, all of its
periodic points lie in the interaction region. This is verified in
Fig.~\ref{fig:p7_man} which  shows an approximation to the $5/7$
manifold, the circles represent the periodic points of the orbit.  

The P7 orbit is also responsible for the length scaling between levels
in Figs.~\ref{fig:fract_3} by way of its stable manifold.  
In Fig.~\ref{fig:level_lowest} we plot the natural log of
the length of the longest regions in each plot, $\ln(\Delta y_0)$, 
versus the number of bounces to escape, $n$.  The points are well fit
by a straight line with a slope of  $m=-0.4867$.   Since the
set of initial conditions is transverse to the $5/7$ manifold the
scaling relationship in the scattering functions is a direct measure
of the scaling of distances between ``fingers'' of the $5/7$
manifold. This suggests that the scale factor found above may be
related to the Lyapunov exponent of the $5/7$ orbit.  The Lyapunov
exponent is found to be $\lambda \approx 0.42$.  The details of this
relationship will be  investigated elsewhere.  Figure \ref{fig:man_sketch} 
shows a cartoon sketch  of the stable manifold, $W_s$, periodic point marked
 $P$, the line of initial conditions, $I$, and the $\gamma^+_{cr}$
line.
Note that we have represented the spacing between intersections of
 $I$ and $W_s$ as approximately constant when, as we have just seen,
the spacing should decrease exponentially as the periodic point is approached.

Figure \ref{fig:man_sketch} also provides an explanation for the
reflection symmetry between the  adjacent levels of
Figs. \ref{fig:fract_3}.  For instance, to scale level 1 into level 2
we must reverse the orientation of the impact parameter as well as
scale the lengths.  This orientation reversing arises from the fact
that the initial conditions for adjacent levels fall on alternating
sides of the stable manifold.  We see that the line of initial conditions
associated with each level alternates from one side of the manifold to
the other. In terms of the manifold sketched in
Fig.~\ref{fig:man_sketch} mapping the finger labeled ``$0$'' into the
finger labeled ``$1$'' requires reflecting $I$ about their common point
and rescaling. We must also reflect about the line of initial
conditions and scale, however, this transformation cannot affect the
one-dimensional scattering functions.  

Now that we understand the mechanism which produces the largest scale
features of the scattering functions, below we examine the next largest
features.  To this end we looked at a set of pictures similar to
those of Figs. \ref{fig:fract_3} where the enlargements are taken from
the leftmost chaotic region in the level 0 plot ($1.116 \stackrel{<}{\sim} y_0
\stackrel{<}{\sim} 1.118$).  This series of
enlargements looks identical to Figs. \ref{fig:fract_3}, except for
the range of impact parameters involved.  Here, however, we find that
the largest smooth regions in each level are $10$ bounces apart.
The orbit responsible for the scaling in this series is the $7/10$
orbit.  The length scales by $0.45$ between levels while the Lyapunov
exponent of the $7/10$ orbit is $\lambda\approx 0.400$.  The same
reasoning that we used for the $5/7$ orbit applies here; the
characteristics of the unstable $7/10$ manifold are responsible for the
self-similar nature of these structures.  

The process of identifying prominent smooth regions and their adjacent
chaotic regions with certain periodic orbits can be continued.  The
next orbit found in this way is the $8/11$. The pattern that emerges
from this process is that the largest features are 
controlled by the lowest period orbits left in the interaction
region.  For the range of energies for which the $3/4$ orbit is still
in the interaction region this orbit behaves approximately like a
boundary orbit; the very finest features (longest escape time
trajectories) of the scattering function show orbits converging to the
 $3/4$ orbit.   This is only approximately true since there are higher
winding number orbits whose manifolds find their way past the $3/4$
and reach the asymptotic region. The property of the P4 orbit that
distinguishes it as a boundary orbit is the fact that the stable P4
orbit has not yet been destroyed by bifurcations while all of the
other orbits between the P2 and P4 are unstable.  The P4 KAM tori that
remain force the flux of trajectories  into
channels containing the unstable orbits which slow the momentum
diffusion of the trajectory.

Using the approximation that the
$3/4$ orbit presents a phase space boundary we can develop an algorithm
to predict the sequence of orbits responsible for the successively
smaller features of the scattering function. Consider all periodic
orbits whose winding numbers are between the $3/4$ orbit and the next
lower period orbit that falls in the asymptotic region, in the current
example this is the $2/3$ orbit. The orbits are ordered from the lowest
winding number to the highest up to some desired periodicity. If an orbit
has a periodic point in the asymptotic region, then all orbits of lower
winding number are also in the asymptotic region and are excluded. The
remaining orbits are then ordered from the lowest period (largest structures)
to the highest (smaller structures), excluding the boundary orbit $3/4$.  The
sequence for $E=0.2856$ is shown in Table \ref{tbl:orbit_seq} for 
orbits up to P19.    
The location of the cutoff winding number (indicated by the vertical
bar in Table \ref{tbl:orbit_seq}) for a fixed deformation is a
function of energy.  We see that as the energy is increased, orbits are
removed from the sequence and the scaling properties of the scattering
function change.

\section{Conclusions}
\label{sec:conclude}
We have studied chaotic scattering on a quadrupole deformed billiard.
The results obtained have general applicability despite our use of a
specific billiard geometry. 
We showed the existence of preferred chaotic scattering directions.
We have explained the origin and organization of the self-similar
structure of the scattering functions in terms of unstable periodic
orbits of the bound billiard system.  Periodic orbits of the bound
system were found using symmetry line theory and the stable manifolds
of some of the unstable orbits were calculated.

We have shown that the existence of preferred scattering
directions is independent of the choice of initial conditions.  This
is true provided the chosen set of initial conditions intersect the
stable manifolds of periodic orbits in the interaction region.  The
escaping trajectories are restricted to leaving the billiard domain in
the regions of sharpest curvature.  This ``localization'' of the
chaotic scattering trajectories results from the existence and
persistence of large P2 KAM regions in the bound phase space.  This
forces trajectories to be funneled into the two regions 
between the tori which correspond to the the large curvature regions
of the billiard.  The trajectories are also localized in escape angle.
This a consequence of the degree of stretching of the phase 
space enforced by the P2 manifolds.  Increasing the deformations will increase
the stretching of the phase space making the allowed escape angle
range larger (for a fixed energy).  The only requirement for the P2 orbits to
play such a central role is that the billiard wall be concave
everywhere.  The details of the functional form of the shape are not
important as long as integrable billiard geometries (circle, ellipse)
are avoided.

The nearly perfect self-similar structure of the scattering functions
is a result of the fractal structure of the stable manifolds of
periodic orbits in the interaction region.  Each such orbit
contributes to the  structure and scaling of the scattering
functions.  The most influential orbits, in the sense that they are
associated with the scaling of the chaotic regions separated by the
largest smooth regions, are the lowest period orbits that are still in
the interaction region at the chosen scattering energy.  Since the
interaction region is defined by the particle energy, changing the
energy  will change the scattering functions.  However,
for small changes in energy for which the lower period orbits remain
in the interaction region the large scale structure should remain
roughly unchanged.  Only when the lowest period orbits fall in the
asymptotic region will there be a major structural change in the
scattering functions.  
 
We presented a general algorithm for relating the periodic
orbits  to the self-similarity  of scattering functions.  
There are, however, several caveats that apply to this simplistic
picture.  First, we have assumed that the Birkoff orbits, those
that arose from the undeformed circular billiard, are the only
contributors to the structure of the scattering functions.  This view
omits effects due to other orbits such as bifurcations of the Birkoff
orbits.  For the large scale structure that we have examined the
omission seems justified but we expect that the existence of these
orbits should have some noticeable effect on smaller scales.  The
other caveat concerns our ability to sort out the relative sizes of
smooth and chaotic regions.  Beyond the three regions that we have
examined and attributed to the lowest three periodic orbits it is
difficult to determine where to look next.  Nevertheless, the
algorithm still sheds some light on the origins of the structure of the
scattering functions.  With these considerations in mind the algorithm
will also be applicable to higher energy scattering
where the P4 orbit is in the asymptotic region.  The $3/4$ 
orbit then becomes the lower boundary in the same sense that the $1/2$
orbit is a lower boundary in specific example discussed.  The upper
boundary will be the next higher period orbit with KAM tori remaining.  For the 
deformation used throughout this paper ($b=0.15$) that orbit is the
$5/6$.  

We also provided a systematic means of finding the symmetric periodic
orbits of the bound billiard by way of symmetry lines. As long as
$r_s(\theta)$ retains the two spatial symmetries $\operator{R_x}$ and 
 $\operator{R_y}$  the six fundamental symmetry lines given in Table
\ref{tbl:gamma} are generally applicable. From these symmetry lines and the
the application of the map the infinite hierarchy of symmetry 
lines can, in principle, be found.  The intersections of these lines
provide the location in phase space of the periodic points of all the
symmetric periodic orbits. The symmetry line methodology is extremely
powerful and can be automated to make the determination of periodic orbits
particularly easy.   This becomes an important consideration when 
semi-classical methods such as trace formulae are being applied.  This
is an important direction of future research since many of the interesting
physical systems that can be modeled by billiards are mesoscopic systems
lying on the border between the quantum and classical worlds.  

There are other interesting situations which we have left
unaddressed.  If the particle energy is lowered significantly the
corresponding asymptotic region shrinks and eventually the upper
(lower) lobes of Fig.~\ref{fig:scan_b1} will overlap the  lower
(upper) interaction region.  This should represent a significant
change in the scattering dynamics in the sense that some trajectories
escaping the upper (lower) interaction region will be re-injected into
the lower (upper).  The effect will be to increase the lifetime of
some trajectories without introducing an infinitely self similar
structure since they will still be primarily near the P2 manifold and
therefore  attracted to the  asymptotic region.  To put this another
way, the signal will be a lot noisier.  However, the directionality of the
scattering will be just as pronounced since the  KAM islands
of the P2 will still be in effect and the escape angle will
necessarily be close to $\pi/2$.  
 
Finally, changing the deformation parameter, particularly making it smaller
will substantially change the bound phase space and therefore the
scattering.  If there are irrational tori remaining they will impose natural
boundaries to the momentum diffusion of trajectories.  This should
generally produce a simplification in the self-similar patterns of the
scattering functions since there will be an absolute winding number
barrier beyond which the asymptotic region is inaccessible.

\thanks{This work was partially supported by a grant from the National
Science Foundation.   We would like to thank Charles Jaffe for his insightful discussions.


\begin{table}
\begin{tabular}{|c|l|c|}
symmetry line 	& equation  		& domain		\\ \hline
 $\Gamma^x_{0,0}$ & $\theta(p)=0$ 	& $~\forall~ ~ p$ 	\\ \hline
 $\Gamma^x_{0,1}$ & $\theta(p)=\pi $ 	& $~\forall~ ~ p$	\\ \hline
 $\Gamma^y_{0,0}$ & $\theta(p)=-\pi/2$ 	& $~\forall~ ~ p$ 	\\ \hline
 $\Gamma^y_{0,1}$ & $\theta(p)=\pi/2$	& $~\forall~ ~ p$ 	\\ \hline
 $\Gamma^p_0$ 	  & $p(\theta)=0$ 	& $~\forall~ ~ \theta$ 	\\ 
 \hline\hline
 $\Gamma^x_{1,0}$ & 
 $p(\theta)= \frac{r_s^\prime(\theta)\sin(\theta) +
	         	r_s(\theta)\cos(\theta) }{ \sqrt{ r_s^2(\theta) + 
	              {r_s^\prime}^2(\theta)}} $   
		&   $\theta \ge 0 $\\ \hline 
			 $\Gamma^x_{1,1}$ & 
		 $p(\theta)=-\frac{
	         r_s^\prime(\theta)\sin(\theta) +
	         r_s(\theta)\cos(\theta)
	       }{ 
	         \sqrt{r_s^2(\theta) + 
	              {r_s^\prime}^2(\theta)}
	       } $ 
		&  $\theta < 0 $\\ \hline 
	$ \Gamma^y_{1,0}$ & 
		 $p(\theta)= \frac{
	         r_s^\prime(\theta)\cos(\theta) -
	         r_s(\theta)\sin(\theta)}{ 
	       \sqrt{ r_s^2(\theta) + 
	              {r_s^\prime}^2(\theta)}}$  
		&   $0 \le |\theta| < \frac{\pi}{2}$ \\\hline
	$ \Gamma^y_{1,1}$ & 
		$ p(\theta)= -\frac{
	         r_s^\prime(\theta)\cos(\theta) -
	         r_s(\theta)\sin(\theta)
	       }{ 
	       \sqrt{ r_s^2(\theta) + 
	              {r_s^\prime}^2(\theta)}
	       }$
		&  $\frac{\pi}{2} \le |\theta| < \pi$ \\\hline  
	$ \Gamma^p_1 $ & ---  	&  --- \\
\end{tabular}
\caption{\it The symmetry lines $\Gamma_0$ and $\Gamma_1$  and their
branches for the symmetries $\operator{R_x}$, $\operator{R_y}$, and
 $\operator{R_p}$.  The second subscript refers to the different
solutions or branches of Eq.~\ref{eq:sym_B0}.} 
\label{tbl:gamma}
\end{table}

\begin{table}
\[
	\begin{array}{c}
	\frac{2}{3} \\ ~ \\ ~ \\
	\end{array}
	\begin{array}{c}
	\frac{11}{16} \\ ~ \\ ~ \\
	\end{array}
	\left |
	\begin{array}{c}
	\frac{9}{13} \\ \uparrow\\ 4 \\
	\end{array}
	\begin{array}{c}
	\frac{7}{10} \\ \uparrow\\ 2 \\
	\end{array}
	\begin{array}{c}
	\frac{12}{17} \\ \uparrow\\ 6 \\
	\end{array}
	\begin{array}{c}
	\frac{5}{7} \\ \uparrow\\ 1 \\
	\end{array}	
	\begin{array}{c}
	\frac{13}{18} \\ \uparrow\\ 7 \\
	\end{array}
	\begin{array}{c}
	\frac{8}{11} \\ \uparrow\\ 3 \\
	\end{array}	
	\begin{array}{c}
	\frac{11}{15} \\ \uparrow\\ 5 \\
	\end{array}
	\begin{array}{c}
	\frac{14}{19} \\ \uparrow\\ 8 \\
	\end{array}
	\begin{array}{c}
	\frac{3}{4} \\ \uparrow\\ \infty \\
	\end{array}	
	\right .
\]
\caption{\it The lowest period orbits up to P19 with winding numbers
between $2/3$ and $3/4$.  The bottom row of numbers orders them from
lowest to highest period.  The P4 is effectively a boundary orbit and 
is labeled by $\infty$ indicating it as the source of the finest scale
structure.}
\label{tbl:orbit_seq}
\end{table}

\begin{figure} 
\caption{A plot of the potential energy surface given by
Eq.~\ref{eq:step_def} with $b=0.15$ and $V_0=1.0$ .\hspace{\textwidth}}\label{fig:potential}
\end{figure} 
 
\begin{figure}
\caption{Both branches of the ten symmetry lines from $\Gamma^x_0$ to
$\Gamma^x_9$ for $b=0.15$. \hspace{\textwidth}}
\label{fig:sym_lines}
\end{figure} 

\begin{figure} 
\caption{The symmetry lines $\Gamma^x_{0,0}$, $\Gamma^x_{0,1}$,
 $\Gamma^x_{9,0}$, and $\Gamma^x_{9,1}$ for $b=0.15$.  The  orbits
found along $\theta=0$ are labeled by their winding numbers.
 \hspace{\textwidth}} 
\label{fig:sym_x_0_9}
\end{figure} 

\begin{figure} 
\caption{The P9 orbits resulting from Fig. \ref{fig:sym_x_0_9}.
\hspace{\textwidth}} 
\label{fig:p9_orbits}
\end{figure}  

\begin{figure}  
\caption{A blowup of $\Gamma_9\cap\Gamma_0$ around the ``$6/9$''
intersections showing deformation parameters from before the
bifurcation to after the bifurcation. The values of the deformations,
$b_i$ are quoted in the text.  \hspace{\textwidth}} 
\label{fig:p9_bif_close}
\end{figure}   

\begin{figure} 
\caption{Phase space near one of the P3 fixed points (a) for
$b=0.1218$ and (b) $b=0.12195$. \hspace{\textwidth}} 
\label{fig:p9_bif_phase}
\end{figure}  

\begin{figure} 
\caption{The bifurcation diagram for the ``m-bifurcation''
associated with the $2/3$ orbit.\hspace{\textwidth}}
\label{fig:m_bif}
\end{figure} 

\begin{figure} 
\caption{The map featuring some of the remaining  KAM regions around the
stable P2, P4, and P6 orbits as well as some higher period orbits.
Also shown are two P6 island chains around the P2 and two chaotic
trajectories.\hspace{\textwidth}}
\label{fig:tori}
\end{figure} 

\begin{figure} 
\caption{The stable manifolds of the $1/2$ orbit and the $3/4$
orbit along with the  $\gamma^\pm_{crit}$ lines for $E=0.2856$.  Also
shown is the 
set of initial conditions, $I$, corresponding to $y_0>=0$, ${p_y}_0=0$,
and ${p_x}_0 = +\sqrt{2E}$.\hspace{\textwidth}}
\label{fig:2_4mans}
\end{figure} 

\begin{figure} 
\caption{A series of enlargements of escape angle versus impact
parameter for $E=0.2856$ and $b=0.15$. The large hash marks
on the $y_0$-axis indicate the  region enlarged in the next
level. The trajectories are all launched
from the left of the potential region with $p_y=0$ and
$p_x=\sqrt{2E}$. The deformation here is $b=0.15$ and the
energy is $E=0.2856$. The escape angle is the angle
of the momentum vector measured from the
$+x$-axis. }
\label{fig:fract_2}
\end{figure} 

\begin{figure}
\caption{Top: the number of bounces, $n$, before escaping versus the
impact parameter, $y_0$.  Bottom: the escape angle, $\Phi$, as a
function of impact parameter. }
\label{fig:fract_1}
\end{figure} 
 
\begin{figure} 
\caption{The critical angle lines and their images under one iteration
of the map for $b=0.15$ and $E=0.2856$. The shaded regions represent
areas through which  particles that have bounced at least once will
exit. }
\label{fig:scan_b1}
\end{figure} 

\begin{figure} 
\caption{The  escape angle ranges obtained from the shaded lobes
of Fig. \ref{fig:scan_b1} overlaid on the billiard shape for initial conditions above $\gamma^+_{crit}$.\hspace{\textwidth}}
\label{fig:sketch}
\end{figure} 

\begin{figure} 
\caption{The number of trajectories, $N$, binned by  (a) escape angle
 $\Phi$ and (b) exit angle $\Theta$ for all trajectories. The data 
is from $10^6$ initial conditions launched between $-\pi < \theta <
\pi$ and $\pi/2 < \gamma < 2.061596$ corresponding to $E=0.2856$ and
a deformation parameter of $b=0.15$.\hspace{\textwidth}}
\label{fig:with_fly_by}
\end{figure} 

\begin{figure}  
\caption{The number of trajectories, $N$, binned by  (a) escape angle
 $\Phi$ and (b) exit angle $\Theta$ excluding fly by trajectories. The data 
is from the $10^6$ of Fig. \ref{fig:with_fly_by}.\hspace{\textwidth}}
\label{fig:no_fly_by} 
\end{figure} 

\begin{figure}  
\caption{The number of trajectories, $N$, binned by  (a) escape angle
 $\Phi$ and (b) exit angle $\Theta$ including all trajectories with
 more than three bounces. The data 
is from the $10^6$ of Fig. \ref{fig:with_fly_by}.\hspace{\textwidth}}
\label{fig:3_plus} 
\end{figure} 

\begin{figure} 
\caption{The critical angle lines and their images under one
iteration of the {\bf inverse} map for $b=0.15$ and $E=0.2856$. The
shaded regions represent areas which lead to trajectories that bounce
at least once. \hspace{\textwidth}} 
\label{fig:scan_b2} 
\end{figure} 

\begin{figure} 
\caption{Number of bounces, $n$, before escaping as a function of
impact parameter $y_0$.  This series covers the same impact parameter
regions as Fig. \ref{fig:fract_2} The $n$ axis is offset by seven and
truncated to a range of $\Delta n = 20$ at
each enlargement to illuminate the role of the P7 orbit and enhance the
self similarity between enlargement levels. \hspace{\textwidth}} 
\label{fig:fract_3} 
\end{figure} 

\begin{figure} 
\caption{The shortest orbits in each of the levels of
 Fig.~\ref{fig:fract_3}. The trajectories are chosen from the center of
the largest smooth regions in each level.\hspace{\textwidth}} 
\label{fig:fract_3a} 
\end{figure} 

\begin{figure} 
\caption{The $5/7$ orbit that is approached by the orbits in
Fig.~\ref{fig:fract_3a}.\hspace{\textwidth}}  
\label{fig:p7_orbit}
\end{figure} 
 
\begin{figure} 
\caption{An approximation to the $5/7$ manifold.  The periodic
points are indicated by the gray dots.} 
\label{fig:p7_man} 
\end{figure}

\begin{figure} 
\caption{A plot of $ln(\Delta y_0)$ vs. $n$ with the $\Delta y_0$s
are the lengths of the longest region in each of the levels of
Figs.~\ref{fig:fract_3}.   Also shown is a linear fit to this data giving a
slope of $m=-.4867$.\hspace{\textwidth}} 
\label{fig:level_lowest}
\end{figure}

\begin{figure} 
\caption{A cartoon sketch of the stable manifold near a periodic
point in the interaction region (above $\gamma^+_{cr}$).  The line
labeled $I$ represents initial conditions.  The numbers correspond to
the level of enlargement of scattering functions.\hspace{\textwidth}} 
\label{fig:man_sketch}
\end{figure}

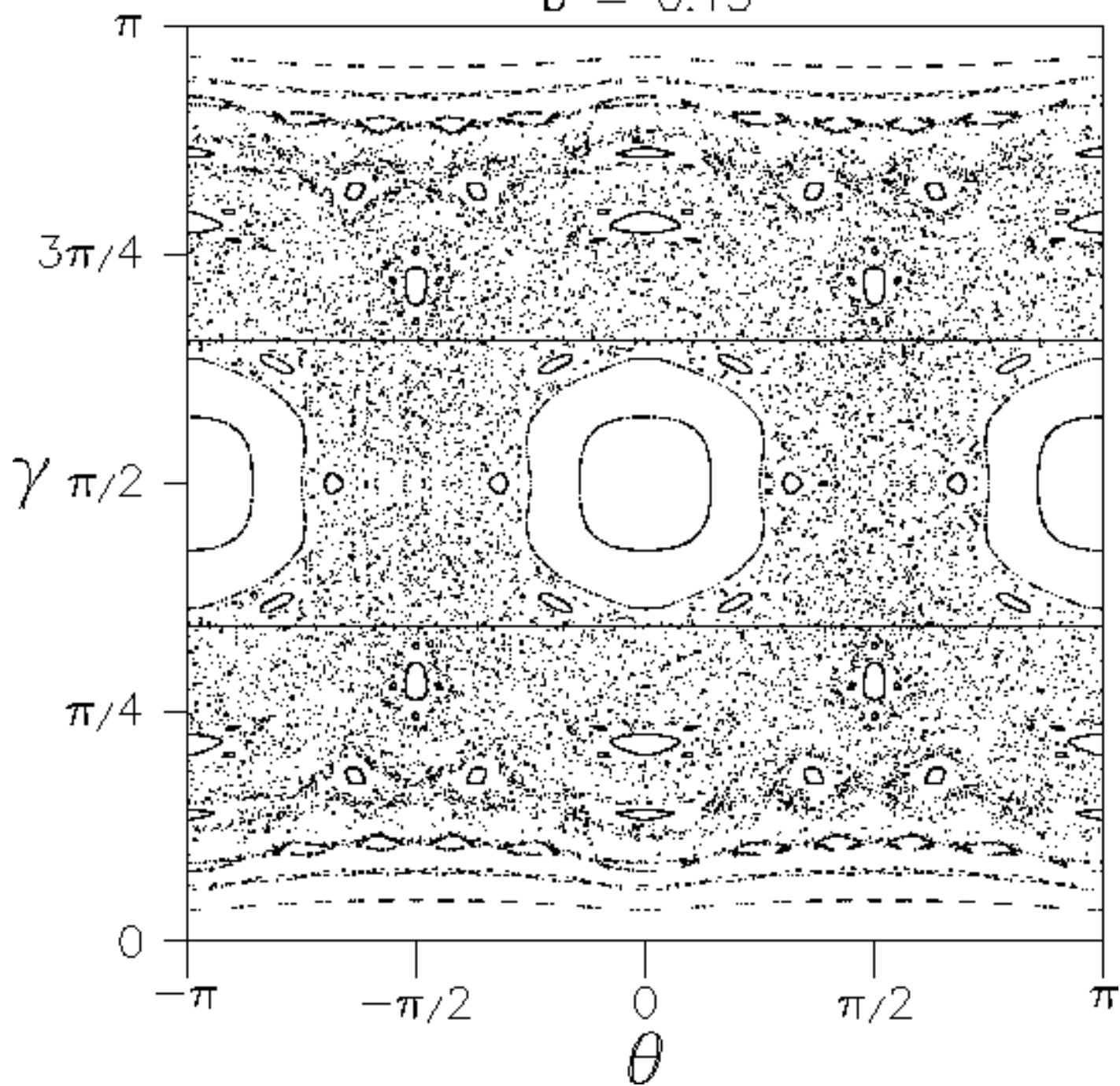
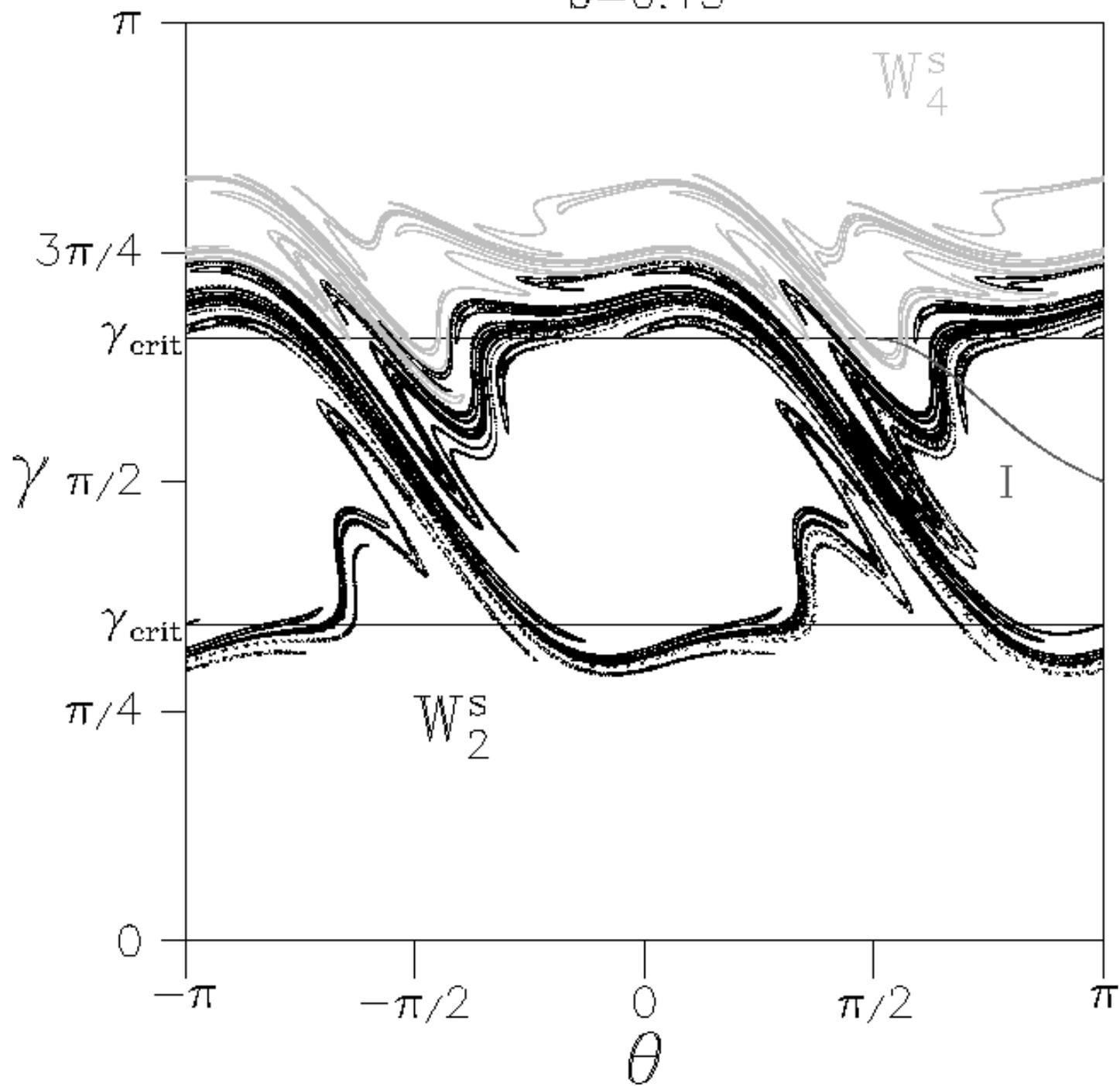
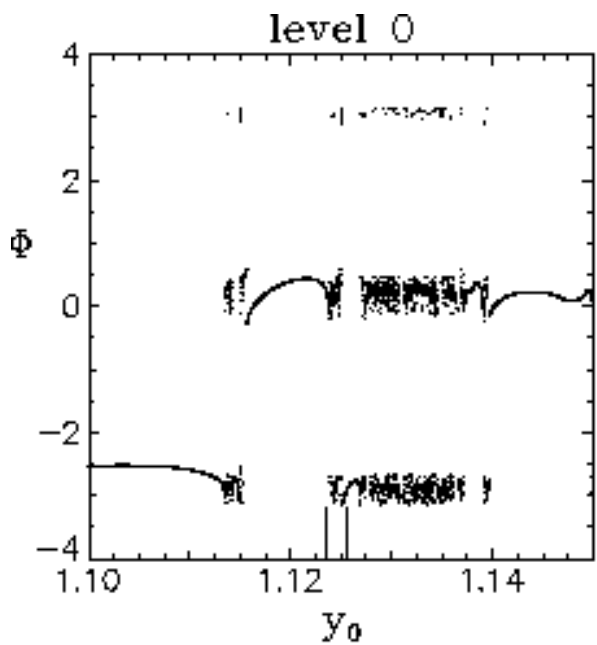
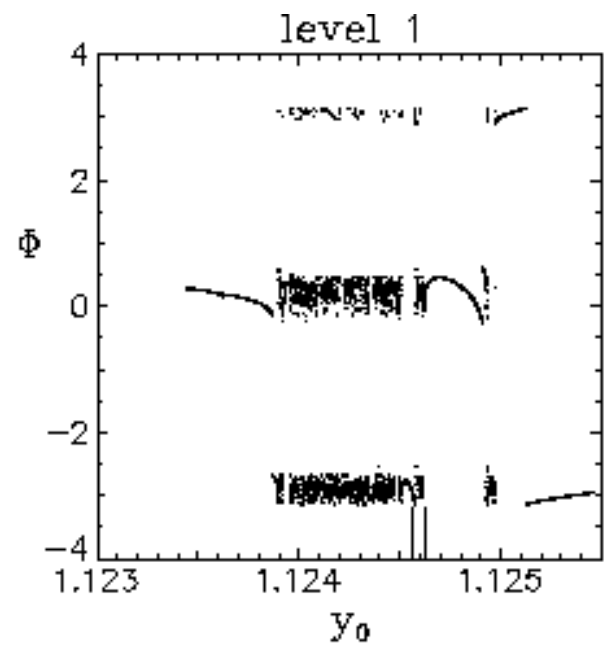
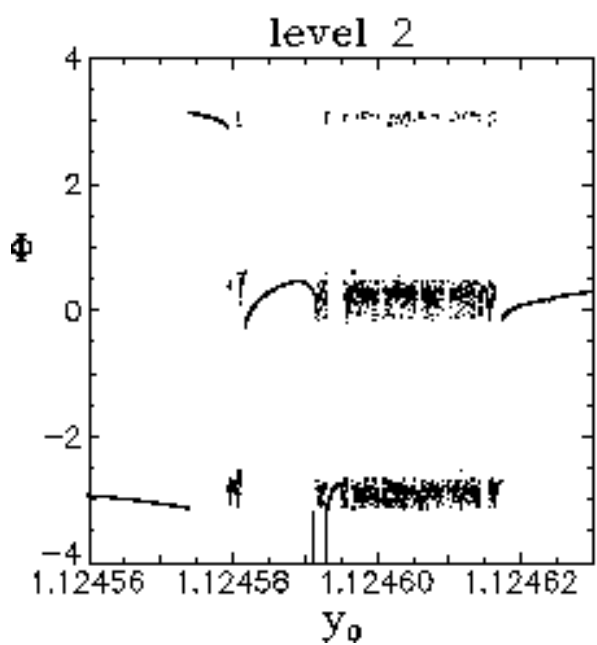
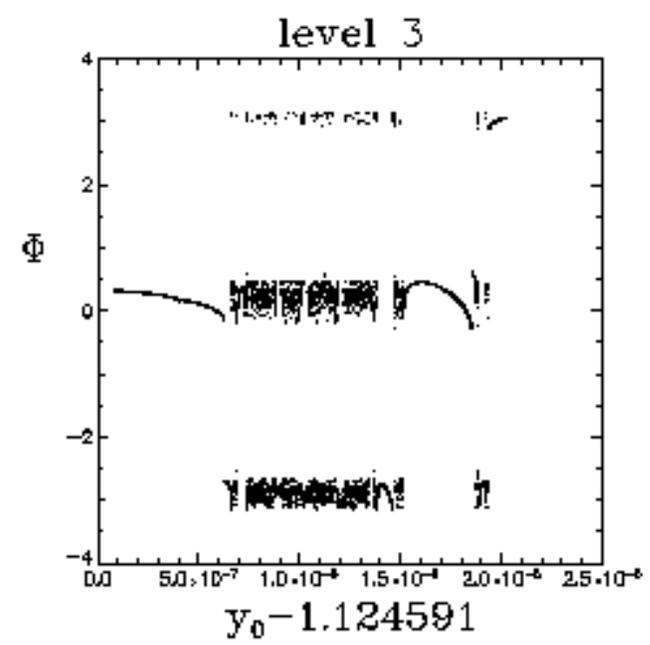
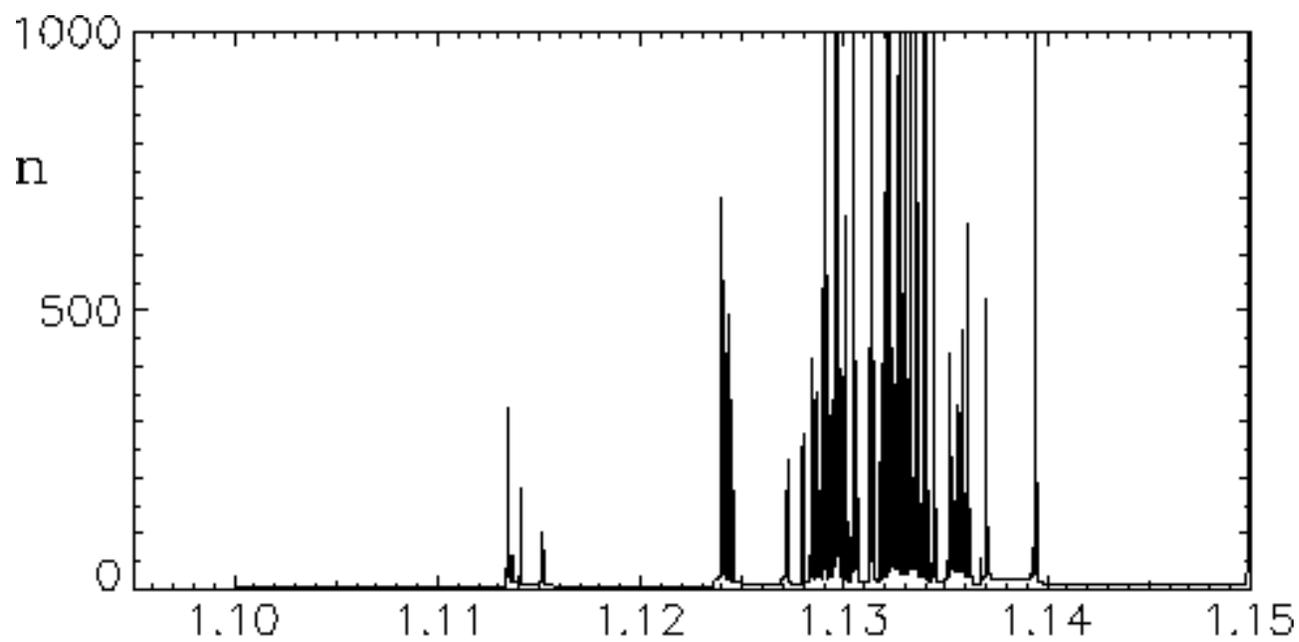
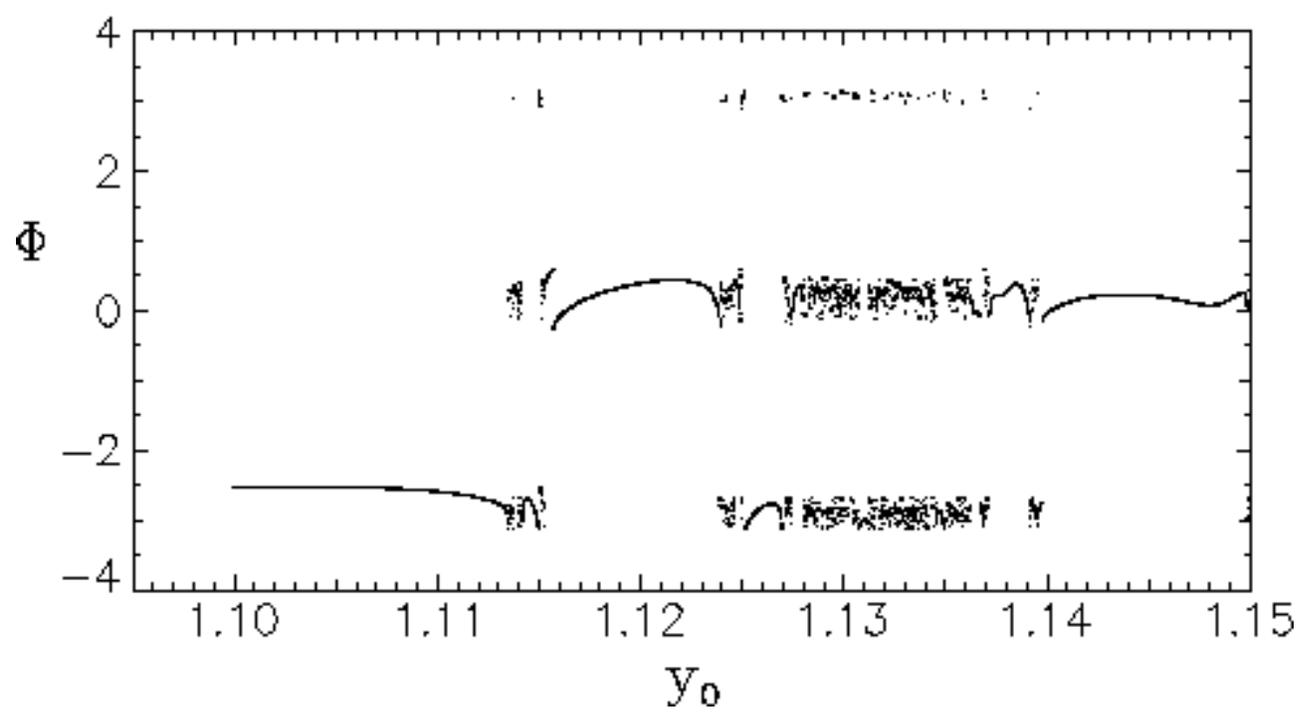
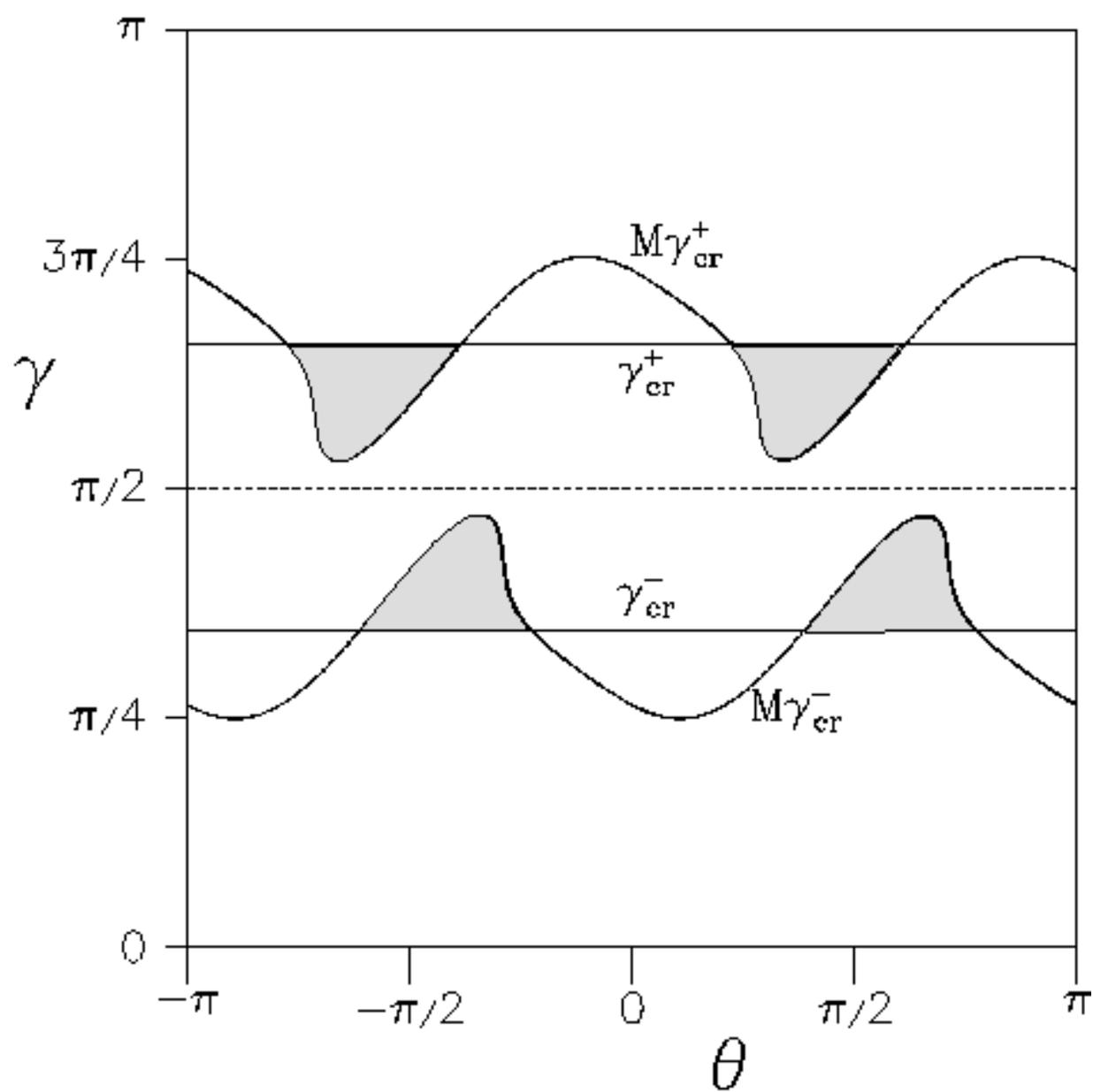
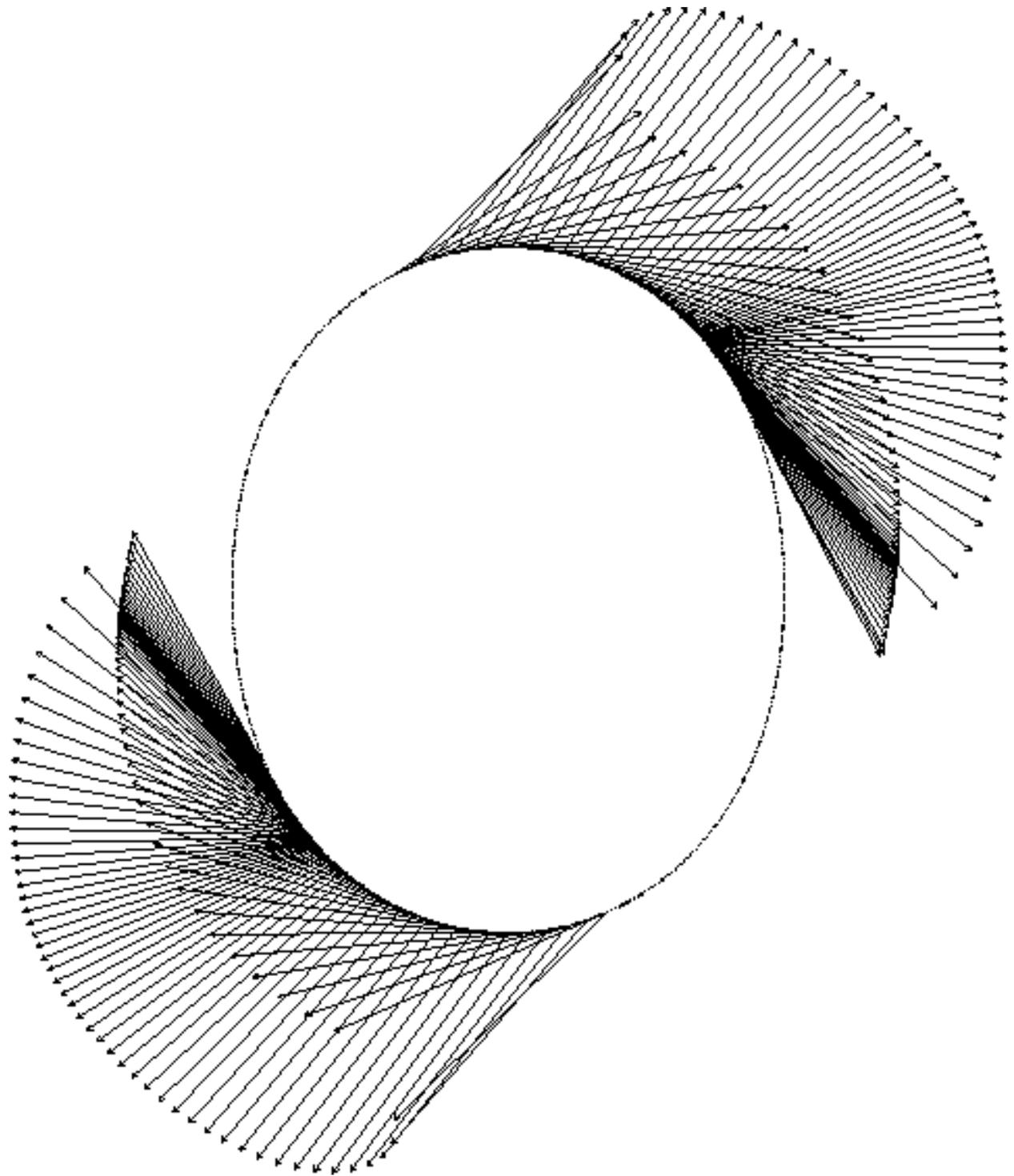
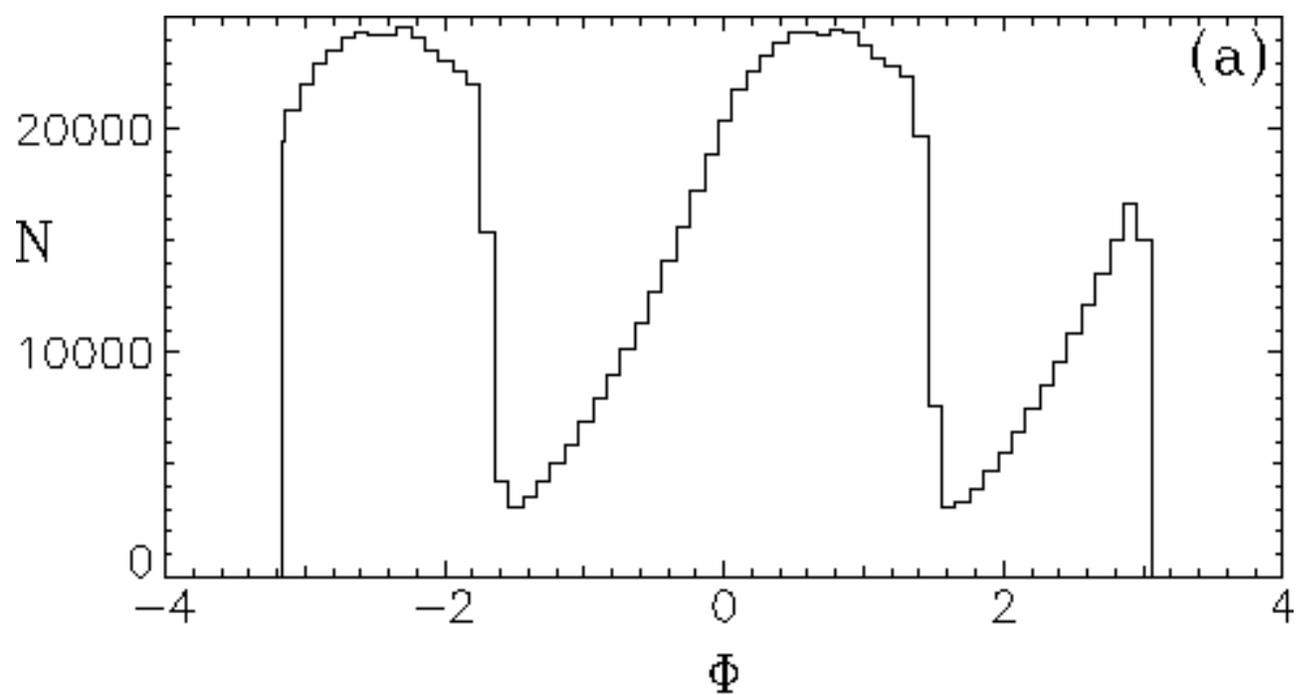
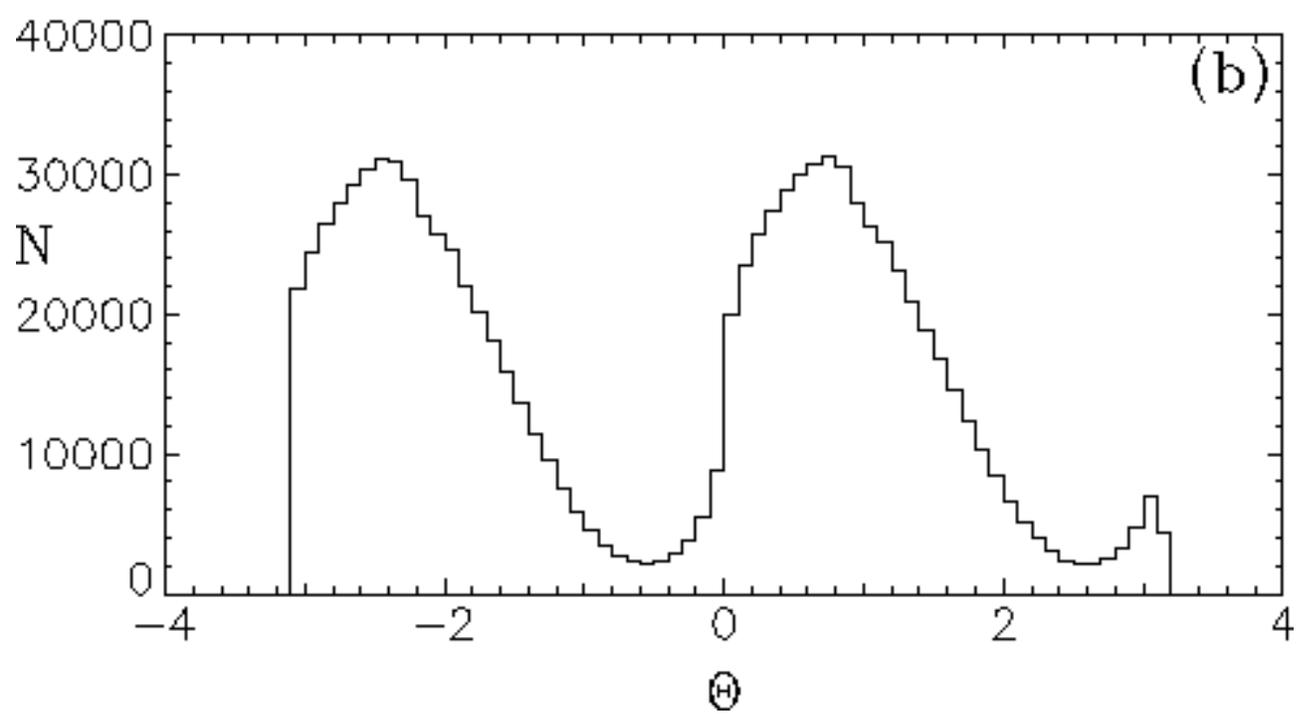
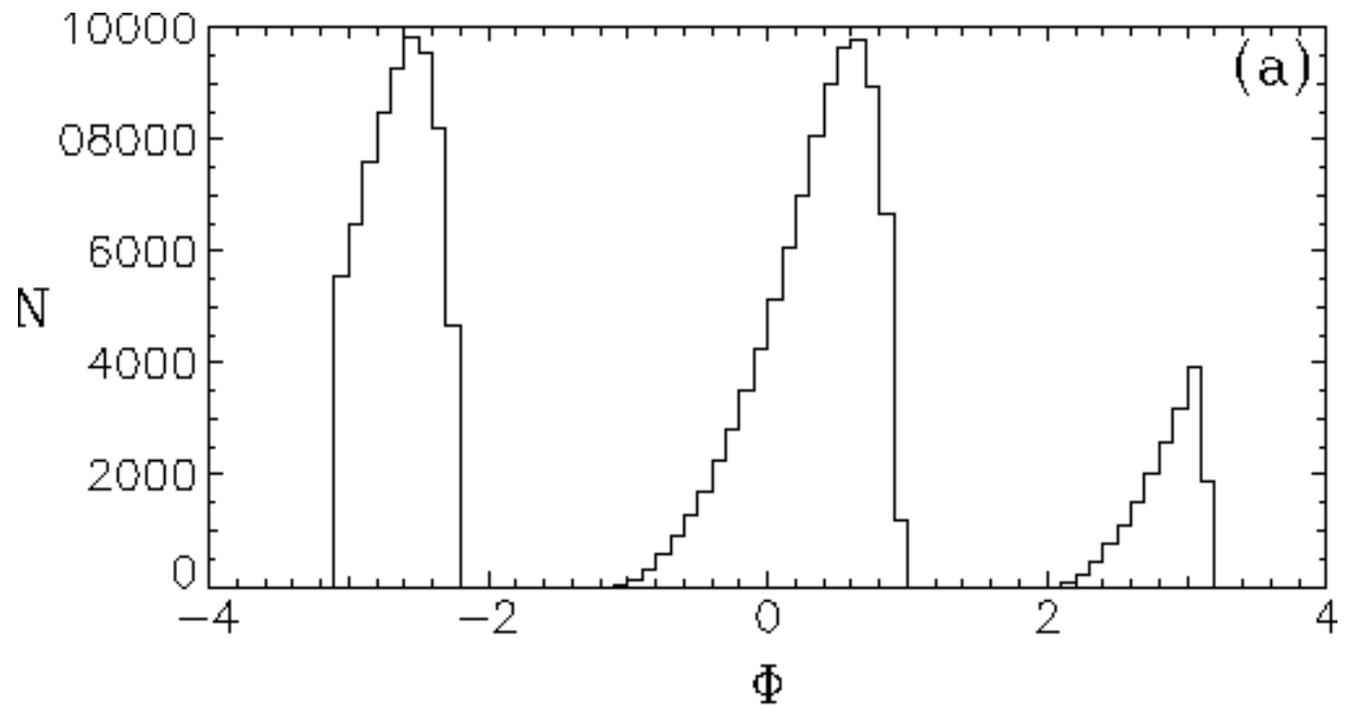
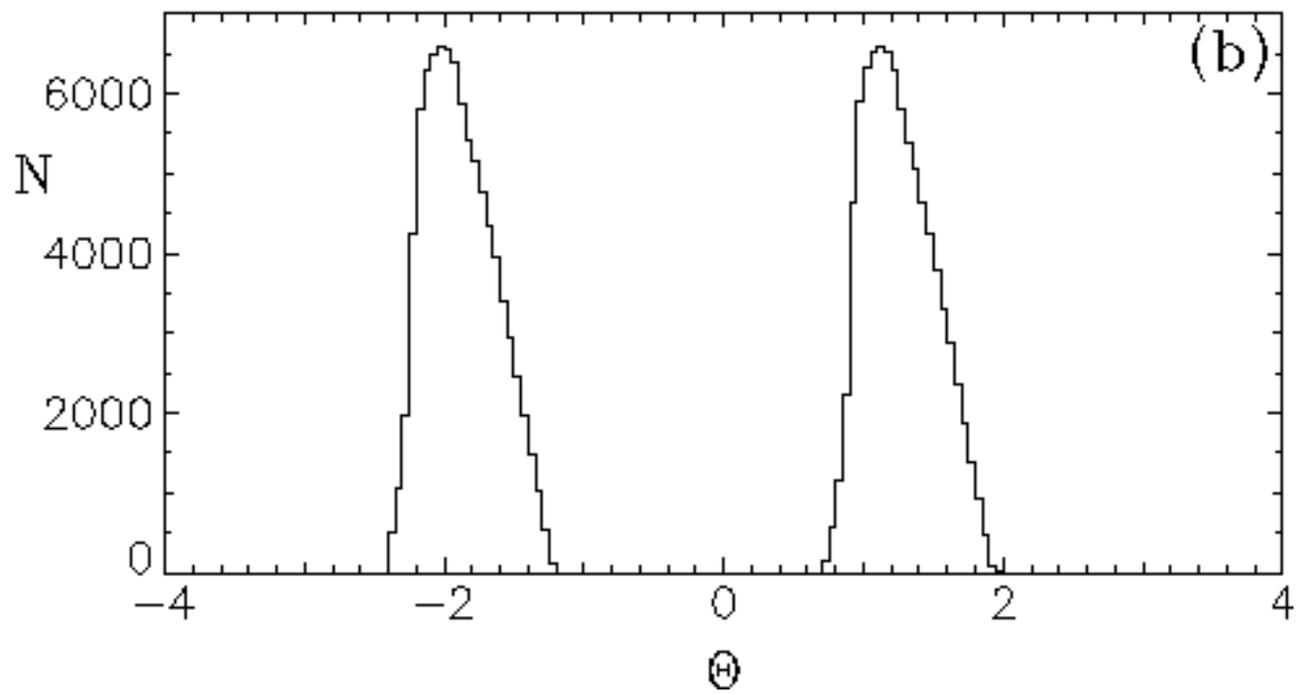
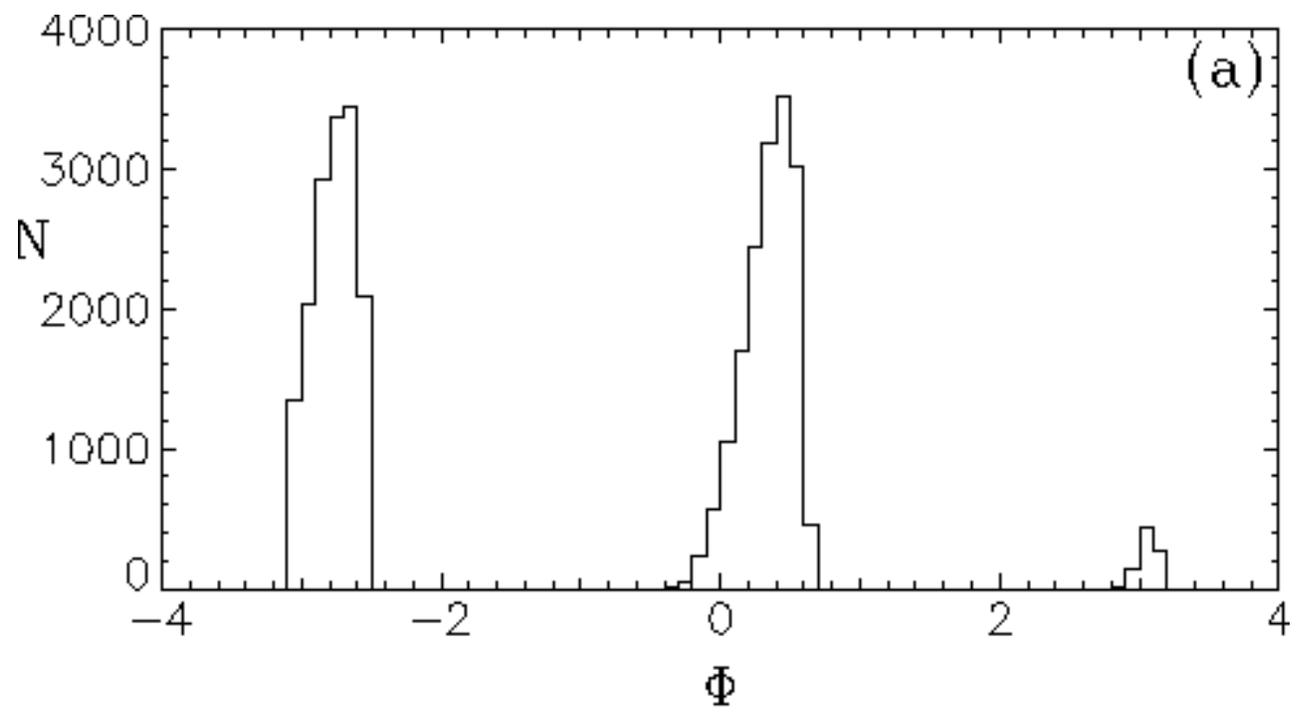
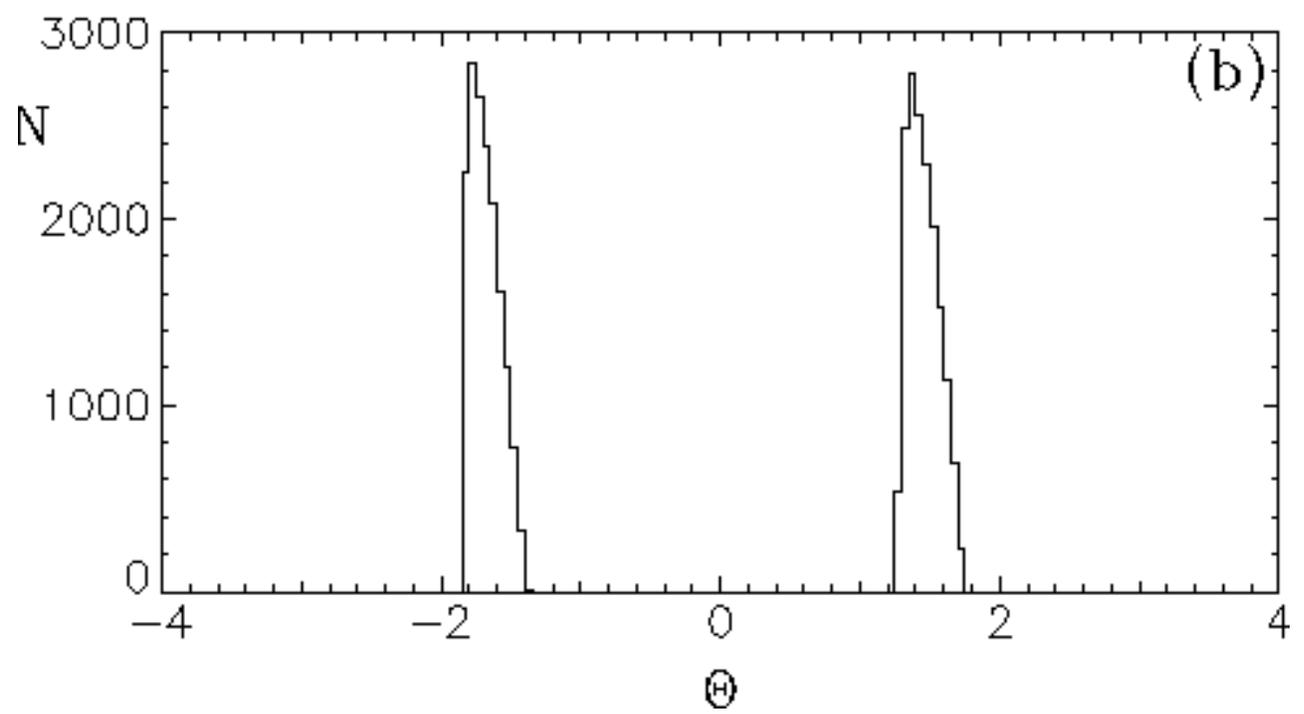
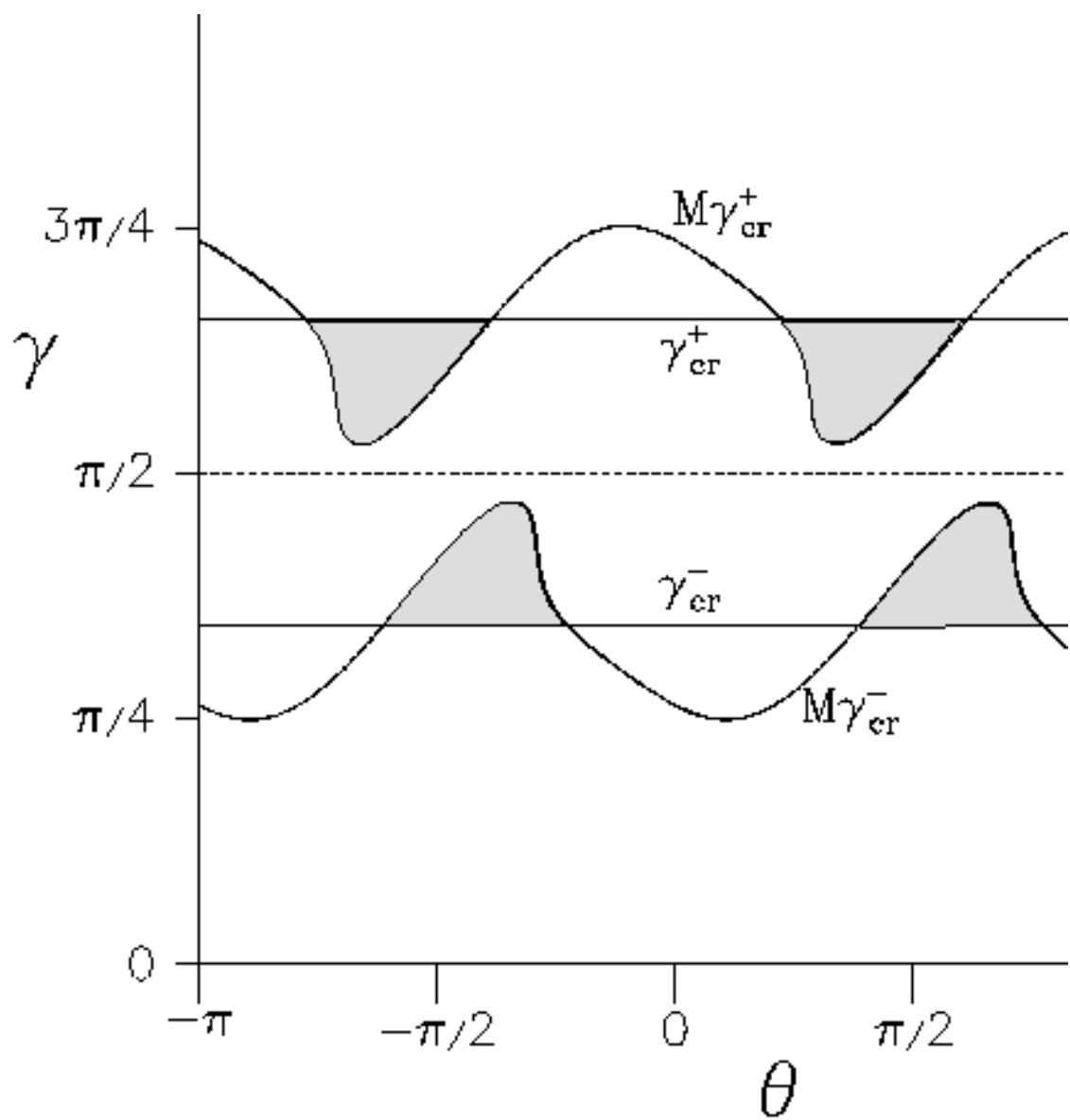
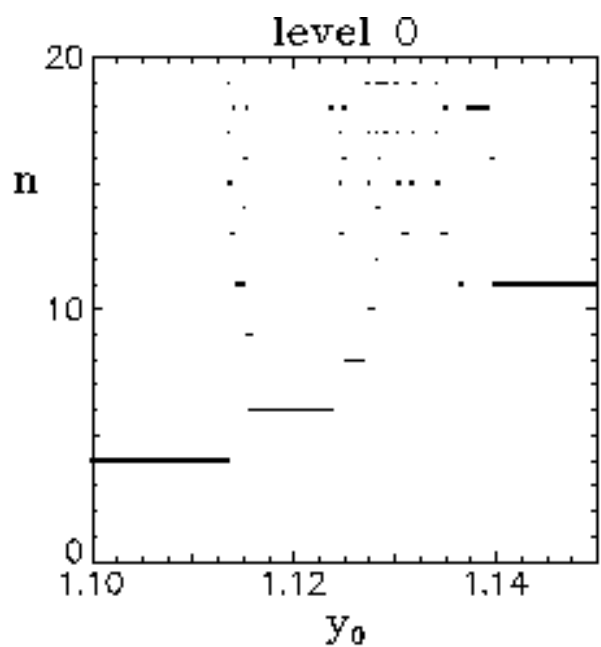
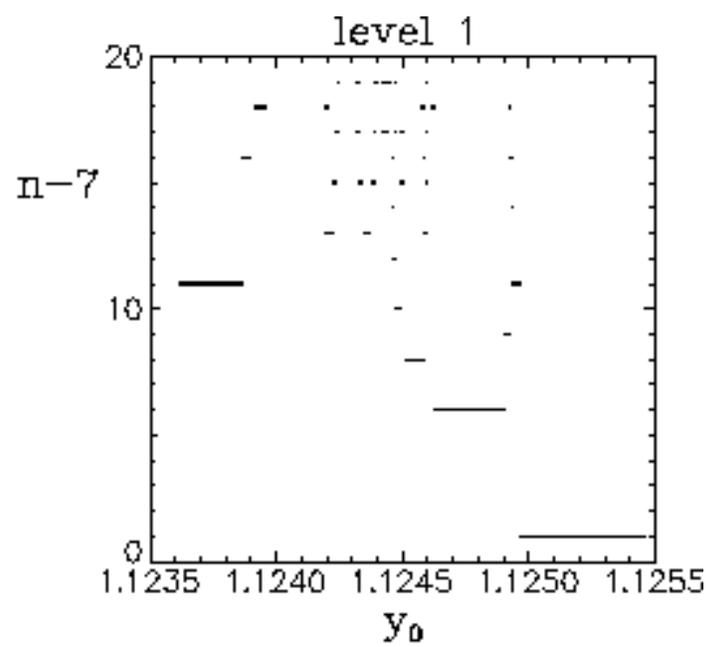
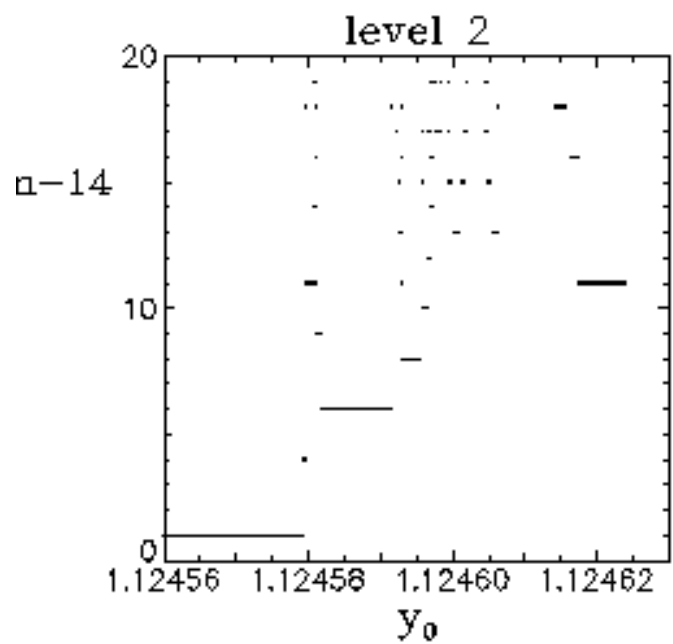
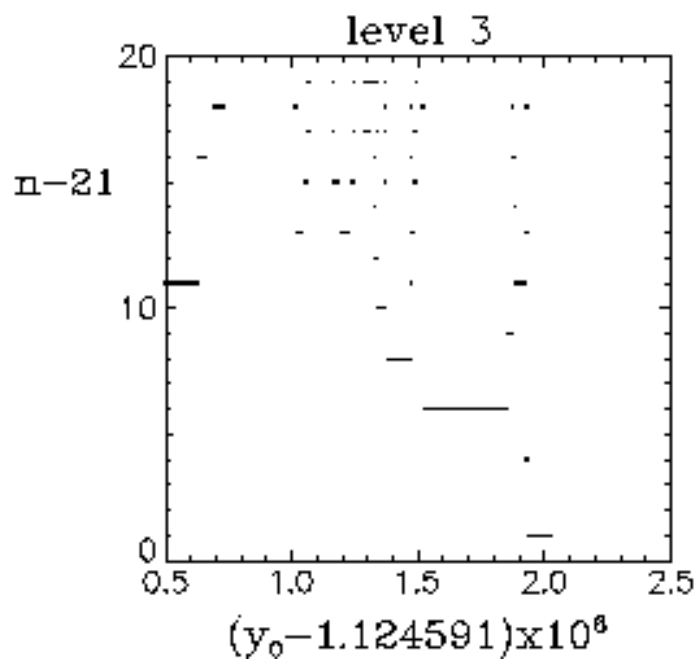
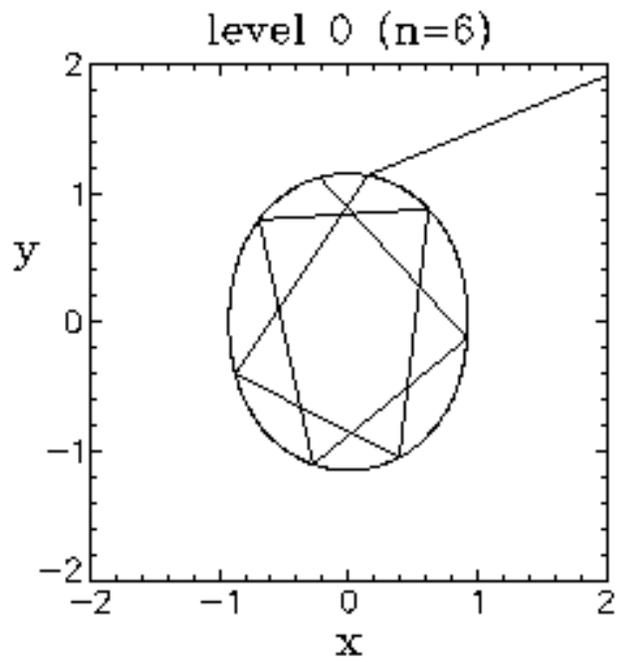
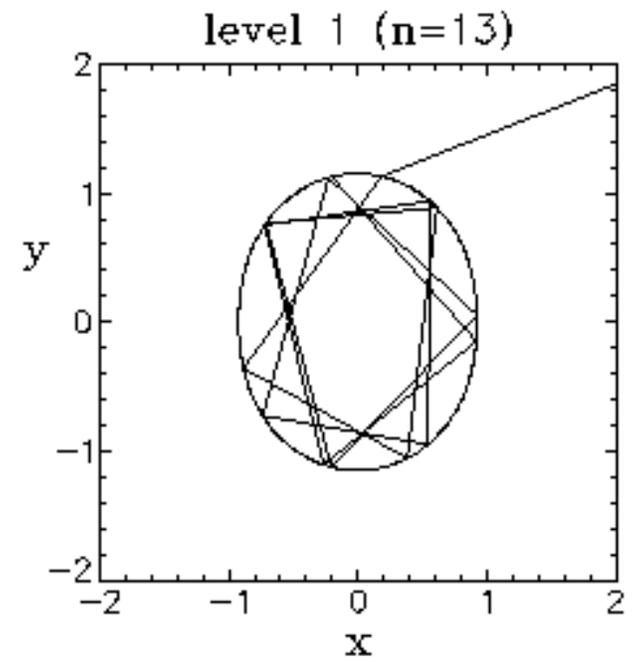
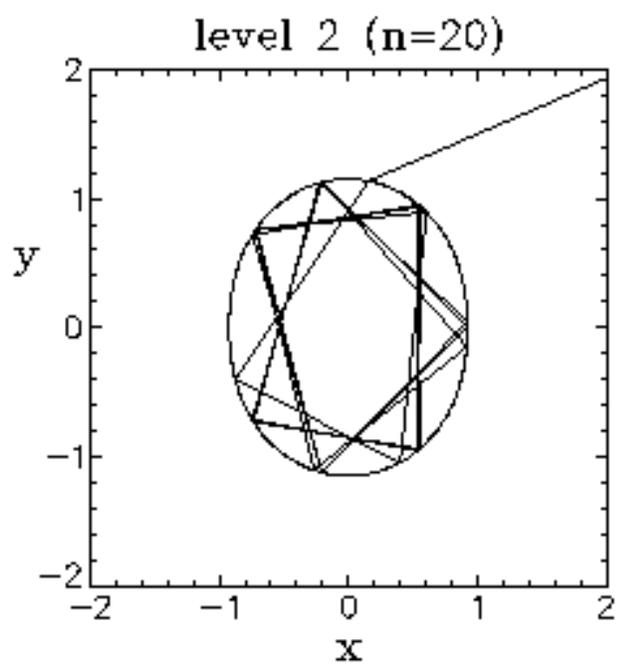
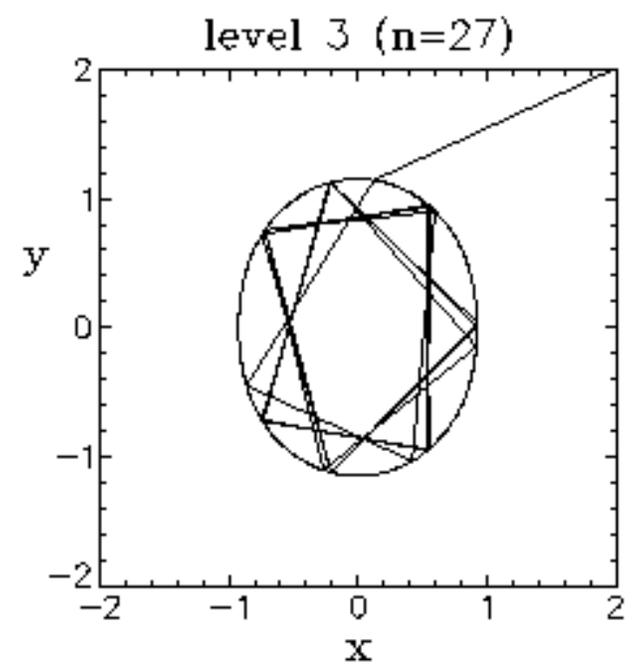
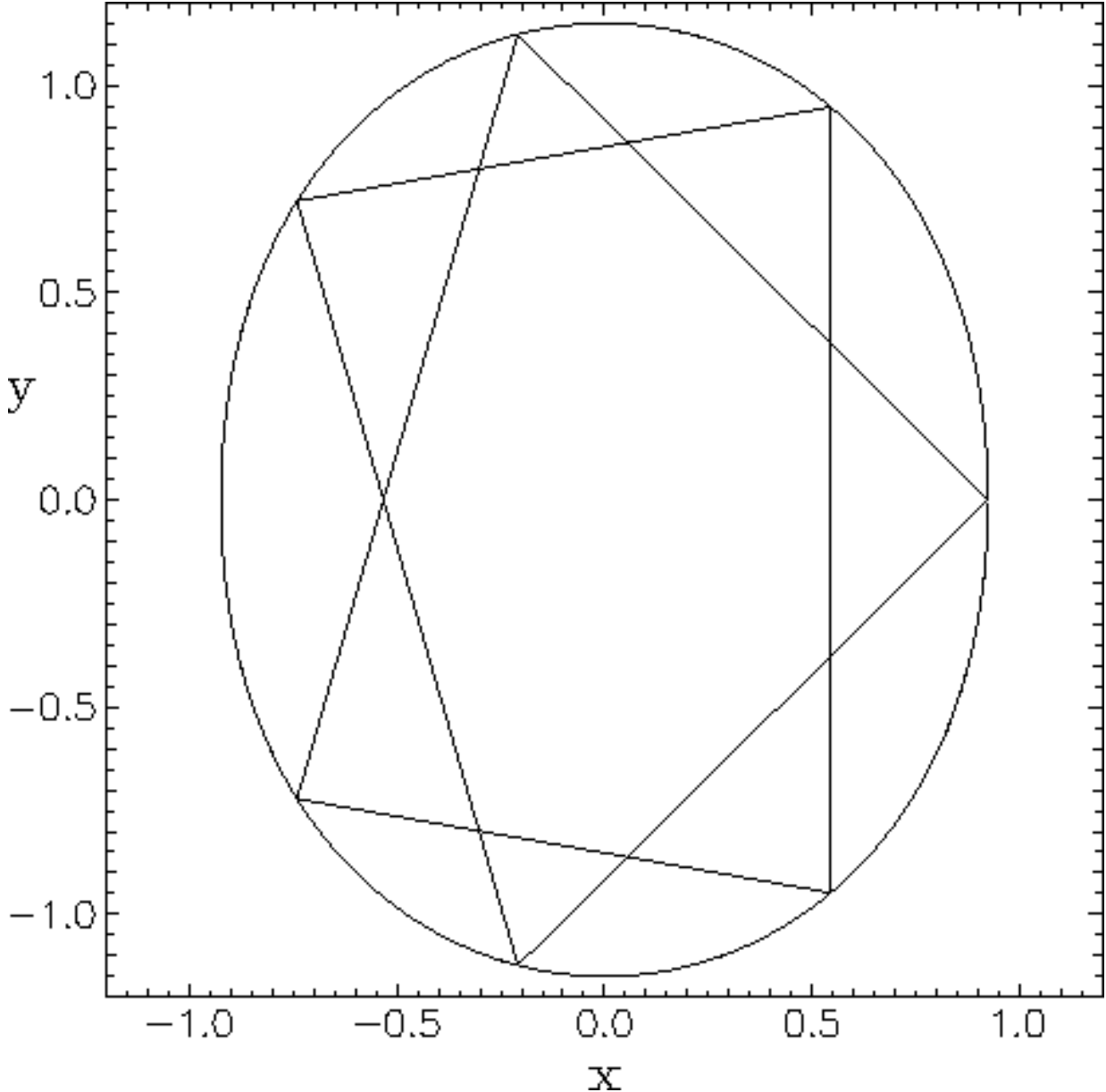
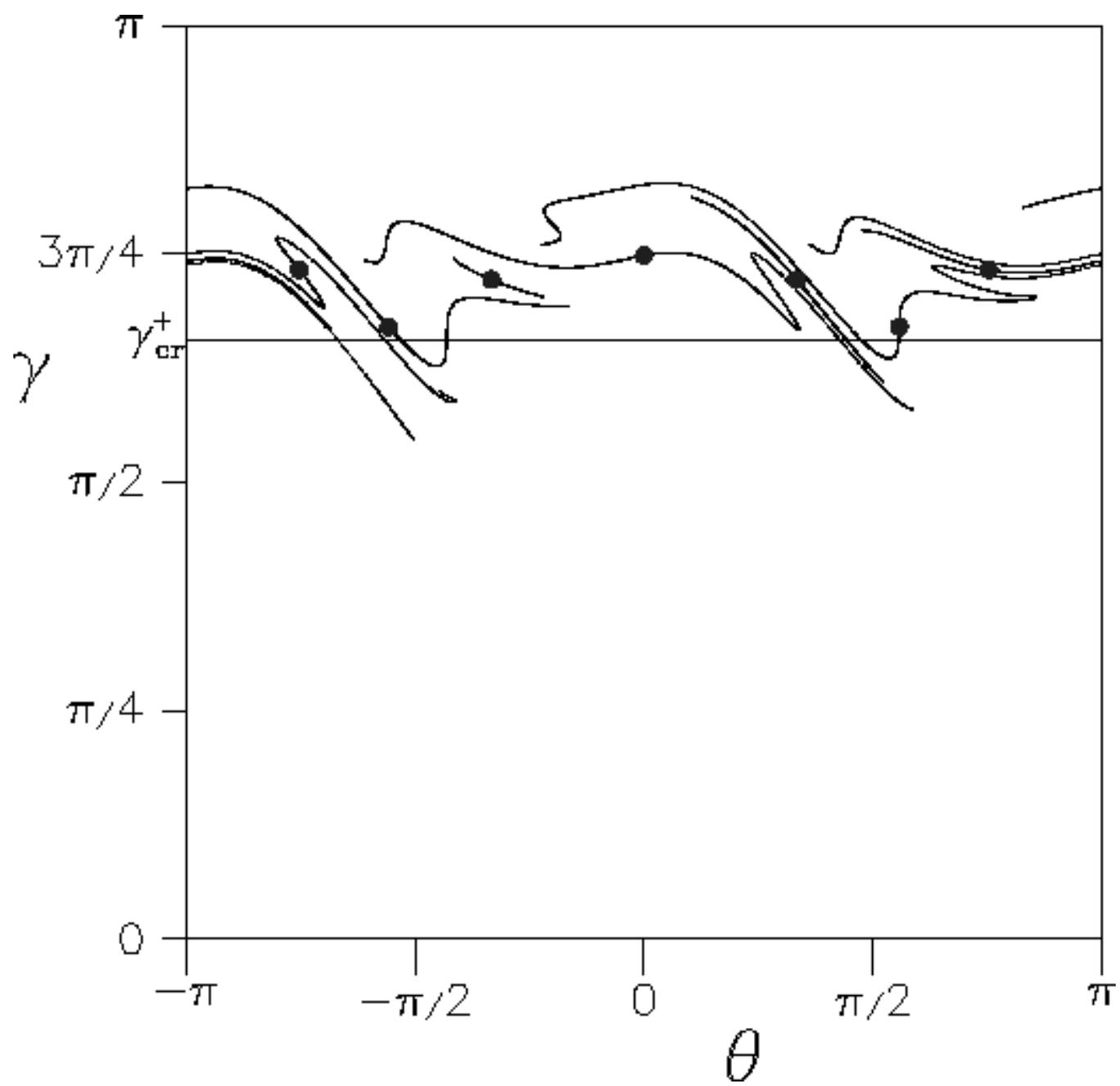
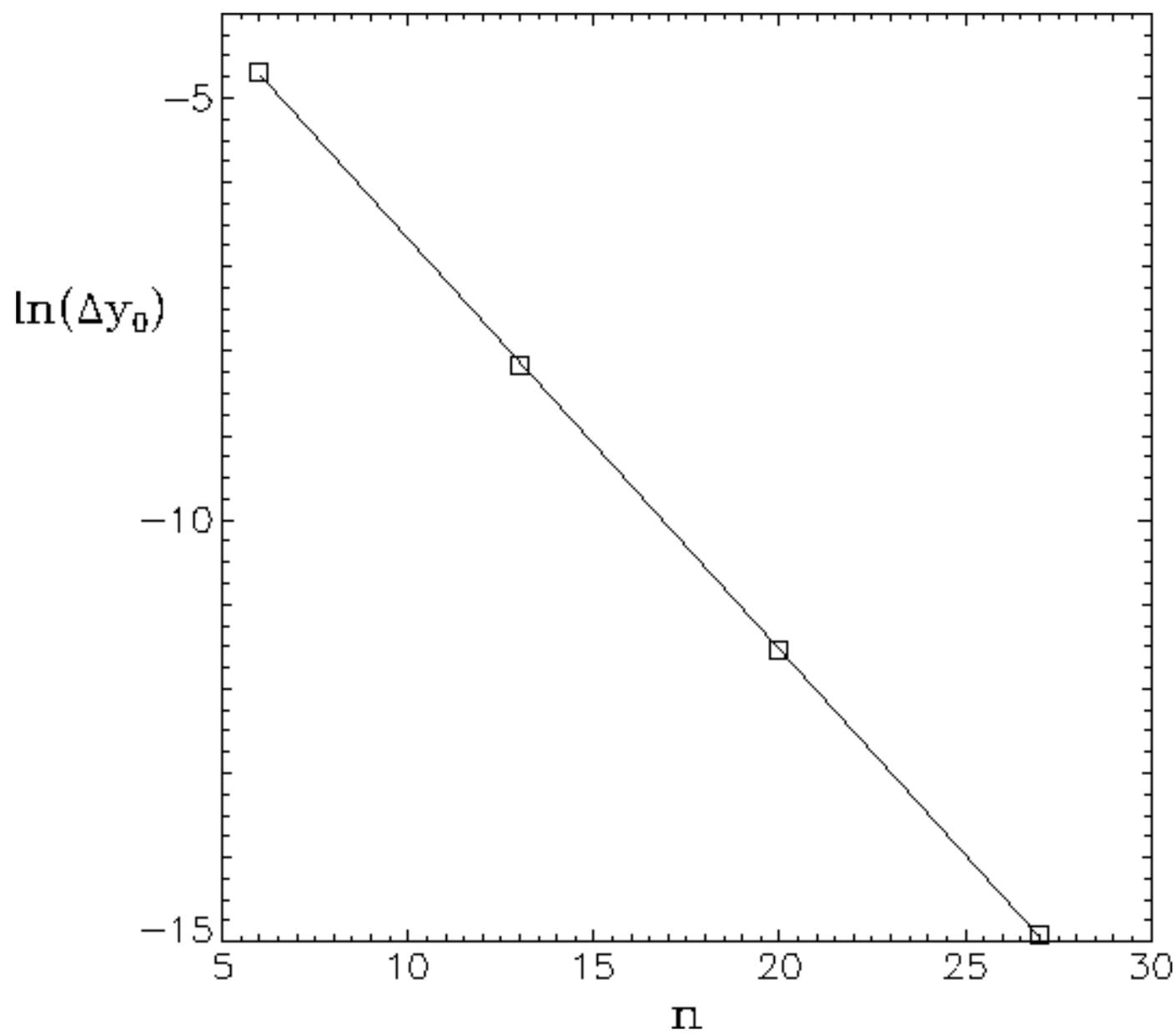
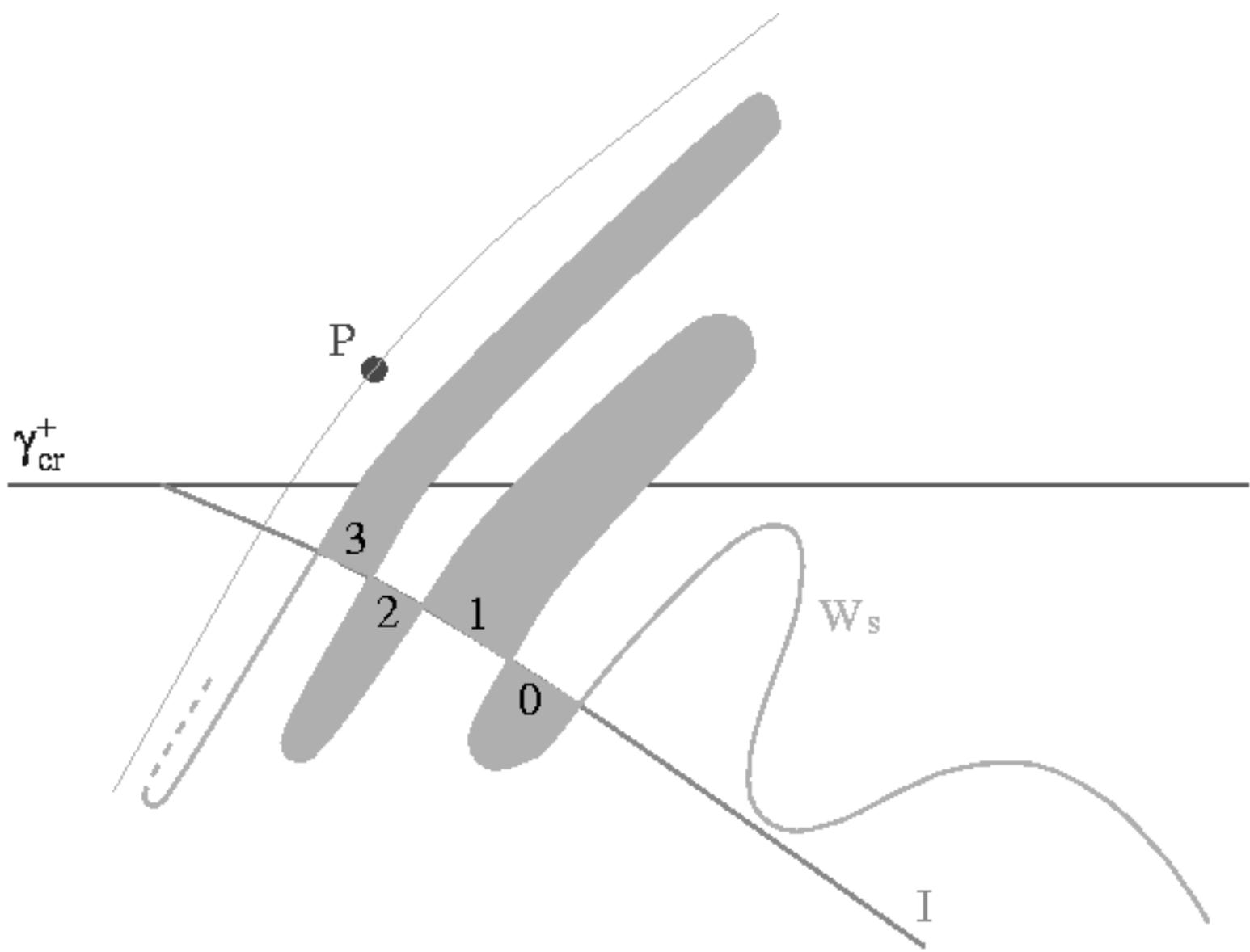
\end{document}